\documentclass[acmsmall,screen,nonacm]{acmart}

\usepackage[english]{babel}

\usepackage{hyphenat}
\usepackage{amsmath}
\usepackage{mathtools}
\usepackage{framed}
\usepackage{stmaryrd}
\usepackage{proof}
\usepackage{listings}
\usepackage{subcaption}
\usepackage{xcolor}
\usepackage{listings}

\usepackage[disable]{todonotes}

\graphicspath{{images/}}
\newcommand{\setZ}{\mathbb{Z}}

\DeclarePairedDelimiter{\denot}{\llbracket}{\rrbracket}
\DeclarePairedDelimiter{\edenot}{\llparenthesis}{\rrparenthesis}

\newcommand{\false}{\text{false}}
\newcommand{\true}{\text{true}}

\newcommand{\eqdef}{\triangleq}

\newcommand{\regr}{\mathsf{r}}
\newcommand{\regc}{\mathsf{c}}
\newcommand{\Reg}{\mathsf{RCmd}}
\newcommand{\ACmd}{\mathsf{ACmd}}

\newcommand{\regplus}{\boxplus}
\newcommand{\kstar}{*}
\newcommand{\Var}{\text{Var}}
\newcommand{\AExp}{\mathsf{AExp}}
\newcommand{\BExp}{\mathsf{BExp}}

\makeatletter
\newsavebox{\@brx}
\newcommand{\llangle}[1][]{\savebox{\@brx}{\(\m@th{#1\langle}\)}%
  \mathopen{\copy\@brx\mkern2mu\kern-1\wd\@brx\usebox{\@brx}}}
\newcommand{\rrangle}[1][]{\savebox{\@brx}{\(\m@th{#1\rangle}\)}%
  \mathclose{\copy\@brx\mkern2mu\kern-1\wd\@brx\usebox{\@brx}}}
\makeatother

\newcommand{\ilexact}[1]{\ensuremath{[#1]}}
\newcommand{\hlexact}[1]{\ensuremath{\{#1\}}}
\newcommand{\silexact}[1]{\ensuremath{\langle #1 \rangle}}

\newcommand{\iltriple}[3]{\ilexact{#1}~#2~\ilexact{#3}}

\newcommand{\iltripleo}[3]{\ilexact{#1}~#2~\ilexact{\oktext{ok: #3}}}
\newcommand{\iltriplee}[3]{\ilexact{#1}~#2~\ilexact{\ertext{er: #3}}}

\newcommand{\hltriple}[3]{\hlexact{#1}~#2~\hlexact{#3}}
\newcommand{\siltriple}[3]{\silexact{#1}~#2~\silexact{#3}}

\newcommand{\silrule}[1]{\ensuremath{\langle \mathsf{#1} \rangle}}
\newcommand{\ilrule}[1]{{\ensuremath{[\mathsf{#1}]}}}
\newcommand{\hlrule}[1]{{\ensuremath{\{\mathsf{#1}\}}}}
\newcommand{\lrule}[1]{\ensuremath{[\mathsf{#1}\rangle}}
\newcommand{\rrule}[1]{\ensuremath{\langle \mathsf{#1}]}}

\definecolor{light-gray}{gray}{0.90}
\newcommand{\ilsiljudgment}[4]{\vcenter{\hbox{\colorbox{light-gray}{\ensuremath{#1}}}}~\vdash~\siltriple{#2}{#3}{#4}}
\newcommand{\ilsiljudgmentv}[4]{\vcenter{\hbox{\colorbox{light-gray}{\ensuremath{#1}}}}~\vDash~\siltriple{#2}{#3}{#4}}
\newcommand{\ilderiv}[4]{\infer{\vdash \iltriple{#2}{#3}{#4}}{#1}}
\newcommand{\uturnderivassertion}[2]{&\quad\ilexact{#1} && \silexact{#2}}
\newcommand{\uturnderivassertionb}[2]{&\quad\ilexact{#1} \quad\rightarrow && \silexact{#2}}
\newcommand{\uturnderivcode}[1]{&\downarrow\quad #1 & \uparrow &}

\newcommand{\turnuderivassertion}[2]{&\quad\silexact{#1} && \ilexact{#2}}
\newcommand{\turnuderivassertiont}[2]{&\quad\silexact{#1} \quad\rightarrow && \ilexact{#2}}
\newcommand{\turnuderivcode}[1]{&\uparrow\quad #1 & \downarrow &}

\newcommand{\judgprq}{\ilsiljudgment{\ilderiv{d}{P}{\regr}{Q}}{P'}{\regr}{Q'}}
\newcommand{\judgvprq}{\ilsiljudgmentv{\ilderiv{d}{P}{\regr}{Q}}{P'}{\regr}{Q'}}

\newcommand{\fwsem}[1]{\denot{#1}}
\newcommand{\bwsem}[1]{\denot{\overleftarrow{#1}}}

\newcommand{\spa}[3]{\fwsem{\code{#1 := #2}}#3}

\newcommand{\Regh}{\mathsf{HRCmd}}
\newcommand{\Cmdh}{\mathsf{HACmd}}

\newcommand{\Val}{\text{Val}}
\newcommand{\Stores}{\text{Store}}
\newcommand{\Heaps}{\text{Heaplet}}

\newcommand{\Asl}{\text{Asl}}
\newcommand{\emp}{\textbf{emp}}
\newcommand{\andsep}{*}

\newcommand{\fv}{\text{fv}}
\newcommand{\dealloc}{\not\mapsto{}}
\newcommand{\fwsemL}[1]{\fwsem{#1}_L \,}
\newcommand{\bwsemL}[1]{\bwsem{#1}_L \,}

\newcommand{\code}[1]{\texttt{#1}}

\newcommand{\svert}{\,\vert\,}

\newcommand{\proofcase}[1]{\noindent \textbf{Case} #1 \newline}
\newcommand{\oktext}[1]{\textcolor[HTML]{004D40}{#1}}
\newcommand{\ertext}[1]{\textcolor[HTML]{D81B60}{#1}}

\newcommand{\okflag}{\oktext{ok}}
\newcommand{\erflag}{\ertext{er}}
\newcommand{\nullt}{\ensuremath{"\backslash{} 0"}}
\newcommand{\nulltc}{"{\char`\\}0"}

\AtBeginDocument{%
	}
\setcopyright{none}
\acmJournal{PACMPL}

\begin{document}
\newtheorem{remark}[theorem]{Remark}

\title{U-Turn: Enhancing Incorrectness Analysis by Reversing Direction}

\author{Flavio Ascari}
\orcid{0000-0003-4624-9752}
\affiliation{\department{Fachbereich Informatik und Informationswissenschaft}\institution{Universität Konstanz}\streetaddress{Universitätsstraße 10}\postcode{78464}\city{Konstanz}\country{Germany}}
\email{flavio.ascari@uni-konstanz.de}
\author{Roberto Bruni}
\orcid{0000-0002-7771-4154}
\affiliation{\department{Dipartimento di Informatica}\institution{Universit\`{a} di Pisa}\streetaddress{Largo B. Pontecorvo 3}\postcode{56127}\city{Pisa}\country{Italy}}
\email{roberto.bruni@unipi.it}
\author{Roberta Gori}
\orcid{0000-0002-7424-9576}
\affiliation{\department{Dipartimento di Informatica}\institution{Universit\`{a} di Pisa}\streetaddress{Largo B. Pontecorvo 3}\postcode{56127}\city{Pisa}\country{Italy}}
\email{roberta.gori@unipi.it}
\author{Azalea Raad}
\orcid{0000-0002-2319-3242}
\affiliation{\department{Department of Computing}\institution{Imperial College London}\streetaddress{180 Queen's Gate}\postcode{SW7 2AZ}\city{London}\country{UK}}
\email{azalea@imperial.ac.uk}

\begin{abstract}

O'Hearn's Incorrectness Logic (IL) has sparked renewed interest in static analyses that aim to detect program errors rather than prove their absence, thereby avoiding false alarms---a critical factor for practical adoption in industrial settings.
As new incorrectness logics emerge to capture diverse error-related properties, a key question arises: 
\emph{can the combination of (in)correctness techniques enhance precision, expressiveness, automation, or scalability?}
Notable frameworks, such as outcome logic, UNTer, local completeness logic, and exact separation logic, unify multiple analyses within a single proof system. 
In this work, we adopt a complementary strategy. Rather than designing a unified logic, we combine IL, which identifies reachable error states, with Sufficient Incorrectness Logic (SIL), which finds input states potentially leading to those errors. As a result, we get a more informative and effective analysis than either logic in isolation.
Rather than naively sequencing them, our key innovation is reusing heuristic choices from the first analysis to steer the second. 
In fact, both IL and SIL rely on under-approximation and thus their automation legitimates heuristics that avoid exhaustive path enumeration (e.g., selective disjunct pruning, loop unrolling). Concretely, we instrument the second logic’s proof rules with derivations coming from the first to inductively guide rule selection and application.
To our knowledge, this is the first rule format enabling such inter-analysis instrumentation.
This combined analysis aids debugging and testing by revealing both reachable errors and their causes, and opens new avenues for embedding incorrectness insights into (a new kind of) scalable, expressive, automated code contracts.

\end{abstract}



\maketitle

\section{Introduction}

Formal methods apply mathematical reasoning to software development, aiming to guarantee correctness, reliability, and security of programs through automated analyses. Over the last decades these techniques have led to several high-profile successes. For example, the Astrée static analyzer uses abstract interpretation to prove the absence of run-time errors in safety-critical C code
~\citep{BlanchetCCFMMMR03},
 Microsoft’s SLAM project (and its Static Driver Verifier tool) applied model checking to millions of lines of Windows driver code, uncovering subtle bugs in API usage~\citep{BallR01},
the CompCert project produced a formally verified C compiler: its machine-checked proof of semantic preservation guarantees that any property proved on the source code holds on the compiled executable~\citep{Leroy09}. Other tools have shown similar impact: the VCC verifier checks annotated concurrent C programs against strong safety and functional properties~\citep{BaudinBBCKKMPPS21}, and the Frama-C platform provides a collaborative, extensible suite of static and deductive analyzers for C, supported by a large academic/industrial community
~\citep{BaudinBBCKKMPPS21}.

 Despite these achievements, the widespread adoption of formal verification in general software development remains limited. 
 A key obstacle is their focus on proving correctness of the program, which often leads to so-called false alarms.
They are warnings produced by the analysis that do not correspond to actual bugs in the code. 
These spurious errors are caused by over-approximations and are particularly frustrating for experienced engineers, who tend to perceive them as distractions rather than helpful insights. This issue becomes even more pronounced in industrial settings, where code is rarely correct on the first attempt. Instead, code development typically follows an iterative process of writing, testing, and refinement. These considerations have prompted a shift in perspective, from \emph{verifying correctness} to \emph{detecting incorrectness}. As a result, there is a growing interest in developing formal methods that are aimed to actively uncover bugs rather than prove their absence. In industrial contexts, under-approximation techniques---such as testing and bounded model checking---are often preferred because they avoid false positives. A notable development in this direction is Incorrectness Logic (IL) \citep{OHearn20}, a program logic specifically designed for bug detection: any error state appearing in the postcondition is guaranteed to be reachable from some input state satisfying the precondition. IL has inspired the creation of practical tools, such as Pulse, which builds on Incorrectness Separation Logic \citep{RaadBDDOV20}, and Pulse-X \citep{LeRVBDO22}.
This foundational work has sparked a new line of research on principled under-approximate approaches to bug detection \citep{MollerOH21,ZilbersteinDS23,AscariBGL25}, as well as on the development of industrial-strength tools for scalable bug finding \citep{DistefanoFLO19,SadowskiAEMJ18}. 

When dealing with real-world applications that must be both effective and scalable, some form of approximation is  unavoidable. This is a direct consequence of Rice’s theorem \cite{Rice53}, which states that any non-trivial semantic property of programs is undecidable.  Therefore, approaches based on under-approximation offer a practical means of scaling analyses, as they allow us to discard part of the information while preserving soundness. As an example, the inference rules of logics for under-approximation can discard disjuncts or bound the number of loops of  iterative commands.
Such abstractions are especially valuable in industrial settings, where incomplete information is the norm and nondeterministic behaviours may emerge simply due to the lack of source code or specifications for external library calls. In these scenarios, under-approximations enable analysis tools to efficiently produce sound results, often at the expense of completeness---that is, the tools may fail to detect all errors---by relying on heuristics that automate the analysis.

\paragraph*{The Problem.}
While early approaches in the literature typically focused on proving either correctness or incorrectness, several successful proposals have since emerged that combine these complementary techniques, resulting in methods that are either more powerful (e.g., \citep{Bradley11, BruniGGR23}) or capable of expressing a broader range of properties (e.g., \citep{ZilbersteinDS23,RaadVO24}). The work that we present here follows this line of thought, firmly grounded in the Aristotelian convinction that ``the whole is greater than the sum of its parts.''
In particular we aim at combining Incorrectness Logic (IL) \citep{OHearn20} with Sufficient Incorrectness Logic (SIL) \citep{AscariBGL25}.
While the goal of IL is to discover reachable error states---such that \emph{any error state in the postcondition can be reached from some input state satisfying the precondition}---the goal of SIL, once the postcondition characterises potential errors, is to identify their sources. In fact, SIL guarantees that \emph{every state in the precondition has an execution leading to an error state in the postcondition}. Both IL and SIL are based on under-approximations; however, while IL can be expressed using a forward semantics, SIL relies on a backward semantics.
IL postconditions expose only true errors, and when paired with the corresponding SIL preconditions, they can be presented to programmers as both reachable errors themselves together with the input states that lead to them, thus aiding the debugging process.
This suggests their combinations is more informative than what each logic can tell individually. 
Indeed, even if sometimes the path conditions leading to an error can be made explicit in the post of IL triples by means of logical variables---variables not appearing in the program and thus never modified during execution---that can keep track of initial values, the next example show that IL triples may not be informative enough to provide the programmer with the input conditions from which to start debugging. 
This is the case, for example,  when the constraints are related to the shape of the heap.
\begin{example}\label{ex:intro-il-issue}
	Consider the following program:
	\[
	\regr \eqdef \code{tmp := [x]; if (tmp == 0) \{ free(x); error() \} }
	\]
	This code fragment reads a value from the pointer $x$ and expects a value different from $0$. However, if it loads $0$, it first deallocates $x$ to reclaim resources before throwing an error. In this case, the path condition leading to the error is that $x$ points to $0$, but this cannot be encoded in the post using logical variables because the heap has changed from the pre to the post, due to the execution of \code{free(x)}.
	In principle, the Incorrectness Separation Logic (ISL) \citep{RaadBDDOV20} triple $\iltriple{x \mapsto 0}{\regr}{\ertext{er: \code{tmp} = 0 \andsep x \dealloc}}$, highlighting the error precondition $x \mapsto 0$, is provable. However, ISL (and IL) validity condition does not give any guarantee on the precondition. Therefore, IL-based proof systems may not infer any meaningful pre.
	For instance, a straightforward application of ISL proof system proves the triple $\iltriple{\code{tmp} = \code{tmp}'}{\regr}{\ertext{er: \code{tmp} = 0 \andsep x \dealloc}}$, which does not contain any reference to $x \mapsto 0$ as the condition leading to the error.
\end{example}

The previous example suggests that the use of SIL, starting from the error postcondition identified through IL, could be useful for generating a warning that also highlights the input that led to the error state, namely the precondition $x \mapsto 0$ in the previous example. 
Our proposal is not to combine IL and SIL in a single proof system where both triples can be derived, but rather to propose a novel proof system that taken a derivation in one logic uses such proof tree to automatically instruct the inference of a triple in the other logic. Next example shows that this can be particularly useful.
\begin{example}\label{ex:intro-sil-issue}
With SIL backward analysis, it is not always possible to determine which execution paths will produce the error. Although this limitation also exists in IL, the highly nondeterministic nature of SIL’s backward semantics makes the problem particularly severe.
	Consider the following program, which inevitably exhibits an erroneous behaviour:
	\[
	\regr \eqdef \code{x := 10; while (x > 0) \{ x-{}- \}; error() }
	\]
	Unfortunately, to detect that $\true$ is a valid SIL pre to an error postcondition, an analyser has to guess that the loop must be executed 10 times. For instance, if the analyser decides to unroll the loop (backward) only 2 times, it will derive the triple $\siltriple{x = 2}{\code{while (x > 0) \{ x-{}- \}}}{\true}$, which will then propagate as $\siltriple{10 = 2}{\code{x := 10}}{x = 2}$ using Hoare's axiom for assignment, which does not expose any cause for the error---in fact, $\false$ is always a valid under-approximation.
\end{example}

To make sure it tracks back some source of errors, SIL analysis should take into account all possible, nondeterministic backward-oriented executions, which is infeasible. Therefore, there's the need for good heuristics to prune the search. 
Another example of the problems raised by high degree of nondeterminism is related to pointer aliasing. While aliasing created during a function execution is easy to detect and track in a forward analysis, it is much harder to infer in a backward analysis; therefore, to find non-trivial preconditions, all possible aliasing must be considered until we reach a point in the code where we can prove they are not admissible. Again, the same issue can happen in a forward analysis, but in practice a function seldom receives two aliased pointers as parameters. On the contrary, it is very common to create some temporary aliases of a pointer for local manipulation, that are discarded before the function returns.
\begin{example}\label{ex:intro-push-back}
	\begin{figure}[t]
	{\scriptsize
		\begin{align*}
			&\silexact{\true \andsep v \mapsto z \andsep z \mapsto - \andsep (x = z \lor x \dealloc{})} \\
			&\quad\code{y := [v];} \\
			&\silexact{\true \andsep v \mapsto - \andsep y \mapsto - \andsep (x = y \lor x \dealloc{})} \\
			&\quad\code{free(y);}\\
			&\silexact{x \dealloc{} \andsep v \mapsto - \andsep \emp \andsep \true} \\
			&\quad\code{y := alloc();} \\
			&\silexact{x \dealloc{} \andsep v \mapsto - \andsep \true} \\
			&\quad\code{[v] := y;} \\
			&\silexact{x \dealloc{} \andsep \true}
		\end{align*}
	}
	\caption{SIL derivation for the reallocation case of \code{push\_back} \citep[Fig.~6]{AscariBGL25}}
	\label{fig:sil-pushback-deriv}
	\end{figure}

	Consider this code fragment that models the reallocation case of C++ \code{push\_back} function (see Example~\ref{ex:uturn-push-back}):
	\[
	\code{y := [v]; free(y); y := alloc(); [v] := y}
	\]
	and the Separation Logic precondition $(v \mapsto x \andsep x \mapsto -)$. Executing the assignment \code{y := [v]} aliases $x$ and $y$, so that the \code{free(y)} deallocates the pointer $x$. It is easy to find this information in a forward analysis at the previous line, where $y$ gets assigned the value pointed by $v$ that is exactly $x$, and find that at the end $x$ is deallocated. On the contrary, if we start from the error postcondition $(x \dealloc{} \andsep \true)$ with a backward analysis, this information is not known until we get to the caller of this code fragment, and therefore we have to consider both possibilities in the pre. This can be seen in the SIL derivation in Fig.~\ref{fig:sil-pushback-deriv} (first presented in \citet{AscariBGL25}), that must account for both cases, whether they are aliased ($x = z$) or they are not ($z \dealloc{} \andsep x \dealloc{}$).
\end{example}

The idea of combining IL and SIL analyses has already appeared in the literature. Notably, in \citet{RaadVO24}, the authors point out the importance of both forward and backward under-approximation information, and exploit it to reason about termination, introducing the new UNTer proof system. UNTer logics effectively proves triples that are valid for both IL and SIL. When turning to the implementation, they realize that Pulse (an industrial-strength automated tool in use at Meta) \emph{already} implemented an analysis that computed triples valid both in IL and SIL (albeit without realizing it explicitly), demonstrating the strength and impact of this combined approach.
However, integrating forward and backward reasoning into a single proof system presents certain challenges.

Designing a proof system that supports both directions crucially relies on formulating appropriate axioms for atomic commands. For instance, while in SIL both classical axioms for assignment used in Hoare logic---Hoare’s backward substitution rule~\citep{Hoare69} and Floyd’s forward transformer~\citep{Floyd67}---remain valid, ensuring similar validity and completeness in a unified system is non-trivial. The axioms are:
\[
\infer[\hlrule{Hoare}]
{\hltriple{q[a / x]}{\code{x := a}}{q}}
{}
\qquad\qquad
\infer[\hlrule{Floyd}]
{\hltriple{p}{\code{x := a}}{\exists x'. p[x'/x]\wedge x=a[x'/x]}}
{}
\]
\noindent
where $q[a / x]$ denotes the capture-avoiding substitution of all free occurrences of $x$ in $q$ with $a$.
Floyd's forward axiom is also valid in IL, but Hoare's axiom is not~\citep[\S 4]{OHearn20}.
Of course, one natural solution is to consider Floyd’s forward axiom, which is valid for both IL and SIL triples. However, this raises an important question: is this the most general axiom we can design for assignments? Could it be that the most general axiom is neither of the previously proposed ones, which were tailored specifically for forward or backward reasoning?

Another key question is: which direction should be prioritized? In other words, should we begin with the precondition and attempt to infer the appropriate postcondition in a forward style (as done in IL and UNTer), or should we start from the postcondition and infer the corresponding precondition in a backward style (as done in SIL)?
In the case of assignment, Floyd’s axiom would be the natural candidate for the first approach and Hoare's axiom for the second.
In a combined proof system such as UNTer’s, proving a triple usually requires the user to make an informed guess for the appropriate pre- and postconditions to use. In fact, it is not the case that for any precondition $p$ (respectively, any postcondition $q$) we can find a corresponding triple that is valid in both IL and SIL. This aspect can be particularly challenging, especially when reasoning about unknown or partially known code, as the user must provide valid triples even in the absence of full information.

\paragraph*{Contribution.}
In response to the above questions, we make two key contributions.

First, we address the problem of formulating axioms for atomic commands that ensure derivation of all and only triples that are valid in both IL and SIL.
To this aim, we propose an \emph{axiom schema} for atomic commands that is sound and complete. From this schema, we derive axioms for atomic commands that are correct by construction, and we demonstrate how such axioms can be instantiated within the UNTer proof system.
Moreover, our contribution goes further: by introducing this general schema, we establish a methodology that can be systematically applied to any atomic command. As a concrete example, we consider non-deterministic assignment---a command not previously supported in UNTer---and immediately derive a new sound and complete axiom for it.

Our second contribution addresses the direction of the analysis. Instead of proposing a single, combined proof system that naturally favours one or the other direction, we suggest a ``smart sequential composition'' of one analysis followed by the other one. 
In particular, in this paper, we instantiate this idea by focusing on the strategy that applies IL followed by SIL---that is, where the results of IL analysis serve as the starting point for SIL. However, the opposite strategy is also possible and briefly outlined in the paper.
As mentioned earlier, IL helps identifying reachable error states, and SIL complements this information by producing inputs and warnings that aid the programmer in debugging their code.
We introduce U-Turn proof system, which allows to follow any IL derivation with a backwards SIL analysis. This combination is not only more informative---since the result satisfies the properties guaranteed by each individual method---but, to the best of our knowledge, it is also the first case where the heuristic exploited by one method is used to guide the application of the other.

\begin{example}
	We briefly revisit the previous examples to show how U-Turn solves their issues.
	
	In Example~\ref{ex:intro-il-issue}, the issue is solved by doing a SIL backward step starting from the error postcondition found by IL. Moreover, this backward step is guided by the forward analysis, that considered the then-branch of the conditional statement, therefore SIL will only analyse it and skip the of the analysis else-branch, finding the desired precondition $\silexact{x \mapsto 0}$.
	
	In Example~\ref{ex:intro-sil-issue}, a first forward step with IL need to unroll the loop until it exits (10 times) to find the postcondition $\ilexact{\ertext{\erflag: x = 0}}$. Taking this information into account, the backward SIL step can unroll the loop for exactly the same number of times to derive $\siltriple{x = 10}{\code{while (x > 0) \{ x-{}- \}}}{\true}$, which will then propagate as $\siltriple{10 = 10}{\code{x := 10}}{x = 10}$ finding the error precondition $\silexact{true}$.
	
	In Example~\ref{ex:intro-push-back}, SIL was already able to infer the error precondition by itself (see~\citep[Ex.~5.2]{AscariBGL25}), but it had to consider both the aliasing and not aliasing of $y$ and $x$. However, the forward IL analysis already has the information that $x$ and $y$ will be aliased. U-Turn is able to transfer this information, forcing SIL to only consider this latter case and drop the other possibility. We show the details of this interaction in Example~\ref{ex:uturn-push-back}. 
\end{example}

\paragraph*{Structure of the Paper.}
The paper is structured as follows.
In Section~\ref{sec:background} we set the notation and introduce relevant concepts from the literature.
In Section~\ref{sec:atoms} we present our first contribution, the forward-backward axiom schema for atomic commands.
In Section~\ref{sec:uturn} we detail our second contribution, the U-Turn proof system.
In Section~\ref{sec:conclusions} we outline possible directions for future works.
Proofs and other technical material can be found in Appendix~\ref{sec:proofs}.

\section{Background}\label{sec:background}

\subsection{Regular Commands}\label{sec:reg}
Following the trend of many other incorrectness logics \citep{OHearn20,RaadBDDOV20,RaadVO24,AscariBGL25} we consider a language of regular commands.
We use standard definitions for arithmetic expressions $\code{a} \in \AExp$ and Boolean expressions $\code{b} \in \BExp$:
\begin{align*}
	\AExp \ni \code{a} ::= \; n \mid x \mid \code{a} + \code{a} \mid \code{a} - \code{a} \mid \code{a} \cdot \code{a} \mid \dots \qquad
	\BExp \ni \code{b} ::= \; \false \mid \lnot \code{b} \mid \code{b} \land \code{b} \mid \code{a} \asymp \code{a}
\end{align*}

\noindent
where $\asymp \in \{ =, \neq, \le, <, \dots \}$ accounts for all standard comparison operators.

The syntax of regular commands $\regr\in\Reg$ is:
\begin{equation}
	\ACmd \ni \regc ::= \; \code{skip} \mid \code{x := a} \mid \code{b?} \mid \code{x := nondet()} \qquad
	\Reg\ni \regr ::= \; \regc \mid \regr;\regr\mid \regr \regplus \regr \mid \regr^\kstar \label{eq:reg-commands-def}
\end{equation}

Note that we include both an explicit nondeterministic assignment \code{x := nondet()} as one of the atomic commands in the language, as well as nondeterministic choice $\regplus$ and iteration $(\cdot)^{\kstar}$.

This formulation accommodates for a standard imperative while-language \citep{winskel93} with the encoding below:
\begin{align*}
	\code{if (b) \{} \code{r}_1 \code{\} else \{} \code{r}_2 \code{\}}& \ \eqdef\ (\code{b?; r}_1) \regplus ((\lnot \code{b})\code{?; r}_2) \\
	\code{while (b) \{r\}} & \ \eqdef\ (\code{b?; r})^\kstar; (\lnot \code{b})\code{?}
\end{align*}

To give a semantics to regular commands, we consider a finite set of variables $\Var$. Let stores $\sigma \in \Sigma \eqdef (\Var \rightarrow \setZ)$ be (total) functions from variables to values. As usual, store update is denoted by $\sigma[x \mapsto v]$. Evaluation of arithmetic and boolean expressions in a store $\sigma$, denoted by $\edenot{\cdot}\sigma$, is standard.
We consider a collecting denotational semantics for regular commands. We define it as a function $\fwsem{\cdot} : \Reg \rightarrow \Sigma \rightarrow \wp(\Sigma)$, which is then lifted to $\fwsem{\cdot} : \Reg \rightarrow \wp(\Sigma) \rightarrow \wp(\Sigma)$ by union. The semantics of atomic commands $\regc \in \ACmd$ and $S \in \wp(\Sigma)$ is defined as follows:
\begin{align*}
	\fwsem{\code{skip}} \sigma &\eqdef \{ \sigma \}
	&\fwsem{\code{x := a}} \sigma &\eqdef \left\lbrace \sigma[x \mapsto \edenot{\code{a}} \sigma] \right\rbrace \\
	\fwsem{\code{b?}} \sigma &\eqdef \{ \sigma \mid \edenot{\code{b}} \sigma = \textbf{tt} \}
	&\fwsem{\code{x := nondet()}} \sigma &\eqdef \{ \sigma[x \mapsto v] \mid v \in \setZ \}
\end{align*}

We then define the semantics of composite regular commands by induction as follows:
\begin{equation}
	\fwsem{\regr_1 ; \regr_2 } \sigma \eqdef \fwsem{\regr_2} (\fwsem{\regr_1} \sigma)
	\qquad
	\fwsem{\regr_1 \regplus \regr_2} \sigma \eqdef \fwsem{\regr_1} \sigma \cup \fwsem{\regr_2} \sigma
	\qquad
	\fwsem{\regr^\kstar} \sigma \eqdef \bigcup_{n \ge 0} \fwsem{\regr}^n \sigma
	\label{eq:fwsem-definition}
\end{equation}

Roughly speaking, given a set of stores $S\subseteq \Sigma$, the collecting forward semantics $\fwsem{\regr} S$ is the set of output states reachable from input states in $S$ by executing $\regr$.

The forward semantics can also be viewed as a binary relation over $\Sigma$, relating a pair of states $(\sigma, \sigma')$ if and only if $\sigma' \in \fwsem{\regr} \sigma$. Following the presentation of SIL \citep[\S 3.1]{AscariBGL25}, we define the backwards semantics $\bwsem{\regr}$ as the function inducing the opposite relation, that is
\begin{equation}
	\sigma \in \bwsem{\regr} \sigma' \iff \sigma' \in \fwsem{\regr} \sigma
	\qquad\text{or, equivalently,}\qquad
	\bwsem{\regr} \sigma' \eqdef \{ \sigma \svert \sigma' \in \fwsem{\regr} \sigma \} .
	\label{eq:bwsem-sigma-sigma'}
\end{equation}
As before, we additively lift the definition of backward semantics to set of states by union.
Roughly speaking, $\bwsem{\regr} S$ is the set of input states that can reach some output state in $S$.\footnote{This was first presented by \citet[\S 5.3]{Hoare78} as the weakest possible precondition calculus. Note that this definition is different from Dijkstra's weakest (liberal) precondition \citep{Dijkstra75}}

\subsection{Assertion Language}\label{sec:assertion-language}
In the paper, we interpret assertions as sets of states. They are described by the following grammar:
\begin{equation*}
	\Asl \ni P, Q ::= P \implies Q \mid \exists x . P \mid \code{b} \mid \fwsem{\regr} P \mid \bwsem{\regr} P
\end{equation*}

Encoding of other logical connectives is standard (eg. $\lnot P \eqdef P \implies \false$, note that $\false$ is part of the syntax of \code{b}). We include in our assertion language constructors for the collecting semantics. While this is theoretically sound, an implementation requires an equivalent closed formula for the semantics, which may or may not be available depending on the command $\regr$.
For instance, there are such closed formulae for both the forward and backward semantics of all atomic commands in our language:
\begin{align*}
	&\spa{x}{a}{P} \equiv \exists v . P[v / x] \land x = a[v / x]
	&&\bwsem{\code{x := a}} Q \equiv Q[a / x]
	\\
	&\fwsem{\code{b?}} P \equiv P \land b
	&&\bwsem{\code{b?}} Q \equiv Q \land b
	\\
	&\fwsem{\code{skip}} P \equiv P
	&&\bwsem{\code{skip}} Q \equiv Q
	\\
	&\fwsem{\code{x := nondet()}} P \equiv \exists x . P
	&&\bwsem{\code{x := nondet()}} Q \equiv \exists x . Q
\end{align*}
where $v$ is a fresh variable (i.e., $v$ does not appear in $P$, $x$ or $a$). Note that the formulae for assignment are precisely the forward transformer of Floyd's axiom for the forward semantics and Hoare's backward substitution for the backward semantics. In the rest of the paper we will often use $\spa{x}{a}{P}$ as a shorthand for $\exists v . P[v / x] \land x = a[v / x]$.

We observe the following relation between forward and backwards semantics of assignments.

\begin{lemma}\label{lmm:assignment-fw-bw}
	Given an assertion $P$, define $Q \eqdef \spa{x}{a}{P}$. Then $P \implies \bwsem{\code{x := a}} Q = Q[a / x]$.
\end{lemma}

\subsection{Incorrectness Logic}

\begin{figure}[t]
	\centering
	\begin{framed}
		\small
		\(
		\begin{array}{cc}
			\infer[\ilrule{assign}]
			{\vdash \iltripleo{\oktext{ok: P}}{\code{x := a}}{\spa{x}{a}{P}}}
			{}
			\quad &
			\infer[\ilrule{assume}]
			{\vdash \iltripleo{\oktext{ok: P}}{\code{b?}}{P \land b}}
			{}
			\\[7.5pt]
			\infer[\ilrule{nondet}]
			{\vdash \iltripleo{\oktext{ok: P}}{\code{x := nondet()}}{\exists x . P}}
			{}
			\quad &
			\infer[\ilrule{skip}]
			{\vdash \iltripleo{\oktext{ok: P}}{\code{skip}}{P}}
			{}
			\\[7.5pt]
			\infer[\ilrule{disj}]
			{\vdash \iltriple{P_1 \lor P_2}{\regr}{Q_1 \lor Q_2}}
			{\vdash \iltriple{P_1}{\regr}{Q_1} & \vdash \iltriple{P_2}{\regr}{Q_2}}
			\quad &
			\infer[\ilrule{cons}]
			{\vdash \iltriple{P}{\regr}{Q}}
			{P \impliedby P' & \vdash \iltriple{P'}{\regr}{Q'}& Q' \impliedby Q}
			\\[7.5pt]
			\infer[\ilrule{seq}]
			{\vdash \iltriple{P}{\regr_1;\regr_2}{Q}}
			{\vdash \iltriple{P}{\regr_1}{R} & \vdash \iltriple{R}{\regr_2}{Q}}
			&
			\infer[\ilrule{er\mbox{-}id}]
			{\vdash \iltriple{\ertext{er: P}}{\regr}{\ertext{er: P}}}
			{}
			\\[7.5pt]
			\infer[\ilrule{choiceL}]
			{\vdash \iltriple{P}{\regr_1 \regplus \regr_2}{Q}}
			{\vdash \iltriple{P}{\regr_1}{Q}}
			&
			\infer[\ilrule{choiceR}]
			{\vdash \iltriple{P}{\regr_1 \regplus \regr_2}{Q}}
			{\vdash \iltriple{P}{\regr_2}{Q}}
			\\[7.5pt]
			\infer[\ilrule{iter0}]
			{\vdash \iltriple{P}{\regr^{\kstar}}{P}}
			{}
			&
			\infer[\ilrule{unroll}]
			{\vdash \iltriple{P}{\regr^{\kstar}}{Q}}
			{\vdash \iltriple{P}{\regr^{\kstar}; \regr}{Q}}
		\end{array}
		\)
	\end{framed}
	\Description{See caption.}
	\caption{Incorrectness Logic rules for regular commands~\cite{OHearn20}}\label{fig:il-rules}
\end{figure}

Incorrectness Logic~(IL) was first introduced in \citet{OHearn20} as a foundation for formal methods aimed to prove program \emph{incorrectness} rather than correctness. The idea of IL is to consider a \emph{subset} or program behaviours rather than a superset. This way, any erroneous behaviour identified by the analysis is proper of the program and not a false alarm induced by the approximation. The ability to consider only subsets of the behaviours by dropping disjuncts and bounded loop unrolling enables scalability at the expense of precision, a worth trade-off in many industrial settings \citep{Godefroid05}.

Formally, the validity of an IL triple $\iltriple{P}{\regr}{Q}$ is defined by the under-approximation condition $\fwsem{\regr} P \supseteq Q$, which is equivalent to
\[
\forall \sigma' \in Q . \exists \sigma \in P . \sigma' \in \fwsem{\regr} \sigma\ .
\]

In other words, any state $\sigma'$ in the postcondition $Q$ is reachable by a real execution of the program starting from some state in $P$, so that all the bugs in $Q$ are reachable.

A hallmark of proof systems based on IL is to tag post, but not pre, with a flag to distinguish between normal and erroneous termination, respectively flagged using the green marker $\okflag{}$ and the red marker $\erflag{}$. In this paper, we instead follow the approach of \citet[\S~6]{BruniGGR21} (see also \citet[Remark~3.9]{AscariBGL25}): instead of attaching flags solely to the postconditions of triples, we enrich the entire state space with them. This is reflected as well in the assertion language, and we assume the semantics of any command acts as the identity on \erflag{}-roneous states, i.e., $\fwsem{\regr} (\ertext{er: \sigma}) = \ertext{er: \sigma}$ for any $\regr \in \Reg$ and $\sigma \in \Sigma$. The benefit of this approach is a more uniform treatment of flags, but it does not introduce any conceptual difference.
This leads us to the modified proof system in Fig.~\ref{fig:il-rules}. Untagged assertions $P, Q$ can contain any disjunction of $\okflag{}$ and $\erflag{}$ states, while tagged assertions $\oktext{ok: P}$ and $\ertext{er: P}$ can only contain states with the specified tag. Axioms for atomic commands are obtained from IL by forcing the pre to only contain $\okflag{}$ states. The only new rule is \ilrule{er\mbox{-}id}, that reflects the identity semantics of any command on $\erflag{}$ states.

As usual, we write $\vDash \iltriple{P}{\regr}{Q}$ for a valid triple and $\vdash \iltriple{P}{\regr}{Q}$ for a provable one.

\begin{theorem}[IL soundness \citep{OHearn20}]
	Any provable IL triple is valid:
	\[
	\vdash \iltriple{P}{\regr}{Q} \implies \vDash \iltriple{P}{\regr}{Q}\ .
	\]
\end{theorem}

\subsection{Sufficient Incorrectness Logic}

\begin{figure}[t]
	\centering
	\begin{framed}
		\small
			\(
			\begin{array}{cc}
				\infer[\silrule{assign}]
				{\vdash \siltriple{\oktext{ok: Q[a / x]}}{\code{x := a}}{\oktext{ok: Q}}}
				{}
				\quad &
				\infer[\silrule{assume}]
				{\vdash \siltriple{\oktext{ok: Q \land b}}{\code{b?}}{\oktext{ok: Q}}}
				{}
				\\[7.5pt]
				\infer[\silrule{nondet}]
				{\vdash \siltriple{\oktext{ok: \exists x . Q}}{\code{x := nondet()}}{\oktext{ok: Q}}}
				{}
				\quad &
				\infer[\silrule{skip}]
				{\vdash \siltriple{\oktext{ok: Q}}{\code{skip}}{\oktext{ok: Q}}}
				{}
				\\[7.5pt]
				\infer[\silrule{disj}]
				{\vdash \siltriple{P_1 \lor P_2}{\regr}{Q_1 \lor Q_2}}
				{\vdash \siltriple{P_1}{\regr}{Q_1} & \vdash \siltriple{P_2}{\regr}{Q_2}}
				\quad &
				\infer[\silrule{cons}]
				{\vdash \siltriple{P}{\regr}{Q}}
				{P \implies P' & \vdash \siltriple{P'}{\regr}{Q'}& Q' \implies Q}
				\\[7.5pt]
				\infer[\silrule{seq}]
				{\vdash \siltriple{P}{\regr_1;\regr_2}{Q}}
				{\vdash \siltriple{P}{\regr_1}{R} & \vdash \siltriple{R}{\regr_2}{Q}}
				&
				\infer[\silrule{er\mbox{-}id}]
				{\vdash \siltriple{\ertext{er: P}}{\regr}{\ertext{er: P}}}
				{}
				\\[7.5pt]
				\infer[\silrule{choiceL}]
				{\vdash \siltriple{P}{\regr_1 \regplus \regr_2}{Q}}
				{\vdash \siltriple{P}{\regr_1}{Q}}
				&
				\infer[\silrule{choiceR}]
				{\vdash \siltriple{P}{\regr_1 \regplus \regr_2}{Q}}
				{\vdash \siltriple{P}{\regr_2}{Q}}
				\\[7.5pt]
				\infer[\silrule{iter0}]
				{\vdash \siltriple{P}{\regr^{\kstar}}{P}}
				{}
				&
				\infer[\silrule{unroll}]
				{\vdash \siltriple{P}{\regr^{\kstar}}{Q}}
				{\vdash \siltriple{P}{\regr^{\kstar}; \regr}{Q}}
			\end{array}
			\)
	\end{framed}
	\Description{See caption.}
	\caption{Sufficient Incorrectness Logic rules for regular commands~\cite{AscariBGL25}}\label{fig:sil-rules}
\end{figure}

While IL only finds true bugs in the post, it does not guarantee anything about the states in the pre. Particularly, thanks to rule \ilrule{cons}, it is always possible to weaken the pre to include states unrelated to the bugs found in the post. This limitation was acknowledged (sometimes less explicitly), e.g., in \citet{LeRVBDO22,ZilbersteinDS23,AscariBGL25}. A proposed solution is a logic that constrains the pre instead of the post, with the meaning that every state in the pre can reach at least one state in the post. Such a logic had multiple names in the literature (Lisbon logic \citep{OHearn20,ZilbersteinDS23}, backwards under-approximate triples \citep{MollerOH21,LeRVBDO22,RaadVO24}, Sufficient Incorrectness Logic \citep{AscariBGL25}).

In this paper, we are interested in combining forward and backwards under-approximation, therefore we take inspiration from the presentation in \citet{AscariBGL25}, but we enrich their rules with error flags to better match the IL rules (these changes were already sketched in \citet[\S~5.6]{AscariBGL25}). The resulting proof system is in Fig.~\ref{fig:sil-rules}. Note that the IL and SIL proof systems have remarkably similar structural rules: the only differences are the rules of consequences and the infinitary rule for iteration (\ilrule{Backwards Variant} in \citet{OHearn20} for IL, \silrule{iter} in \citet{AscariBGL25} for SIL), but the latter is disregarded in this paper. This will allow us to follow derivations in one proof systems using the other one.

Validity of a SIL triple $\siltriple{P}{\regr}{Q}$ is defined by the equation $\bwsem{\regr} Q \supseteq P$, which is equivalent to
\[
\forall \sigma \in P . \exists \sigma' \in Q . \sigma' \in \fwsem{\regr} \sigma\ .
\]
Again, we write $\vdash \siltriple{P}{\regr}{Q}$ for a provable triple and $\vDash \siltriple{P}{\regr}{Q}$ for a valid one.

\begin{theorem}[SIL soundness \citep{AscariBGL25}]
	Any provable SIL triple is valid:
	\[
	\vdash \siltriple{P}{\regr}{Q} \implies \vDash \siltriple{P}{\regr}{Q}\ .
	\]	
\end{theorem}

We conclude observing the following simple facts about valid SIL and IL triples

\begin{lemma}\label{lmm:p-q-valid-empty}
For any regular command $\regr$ and any assertions $P$ and $Q$, it holds:
	\begin{enumerate}
		\item If $\vDash \siltriple{P}{\regr}{Q}$ and $Q = \varnothing$, then $P = \varnothing$
		\item If $\vDash \iltriple{P}{\regr}{Q}$ and $P = \varnothing$, then $Q = \varnothing$
	\end{enumerate}
\end{lemma}

\begin{corollary}\label{lmm:p-q-valid-not-empty}
For any regular command $\regr$ and any assertions $P$ and $Q$, it holds:
	\begin{enumerate}
		\item If $\vDash \siltriple{P}{\regr}{Q}$ and $P \neq \varnothing$, then $Q \neq \varnothing$
		\item If $\vDash \iltriple{P}{\regr}{Q}$ and $Q \neq \varnothing$, then $P \neq \varnothing$
	\end{enumerate}
\end{corollary}

\subsection{Separation Logic}
In this section we give a brief primer on Separation Logic (see, eg., \citet{OHearn19} for an introduction and \citet{OHearnRY01} for a more technical explanation).

First, we augment the program syntax with primitives to operate on the heap. We consider a different set of heap atomic commands $\Cmdh$, and use them to obtain the full language of heap regular commands $\Regh$:
\begin{align*}
	\Cmdh \ni \regc ::= &\; \code{skip} \mid \code{x := a} \mid \code{b?} \mid \code{x := nondet()} \\
	\mid &\; \code{x := alloc()} \mid \code{free(x)} \mid \code{x := [y]} \mid \code{[x] := y} \\
	\Regh \ni \regr ::= &\; \regc \mid \regr;\regr\mid \regr \regplus \regr \mid \regr^\kstar
\end{align*}

Roughly speaking, the semantics of heap regular commands is interpreted over sets of pairs $\Stores \times \Heaps$. A \emph{heaplet} is a partial function $h \in \Heaps = (\setZ \rightharpoonup \Val \uplus \{ \delta \})$, where the input is interpreted as a memory address. Intuitively, a heaplet $h$ only describes a \emph{portion} of the global heap: any location not in the domain of $h$ is unknown (it may be not allocated or belong to a different heaplet); the special value $\delta$ describe a known-to-be-deallocated location.
We use notation $h[l \mapsto v]$ for function update (possibly adding $l$ to the domain of $h$), $[]$ for the empty heaplet (ie. the heaplet with an empty domain) and a list notation $[l \mapsto v]$ as a shorthand for $[][l \mapsto v]$ (ie. the heaplet mapping $l$ to $v$ and undefined anywhere else).

The assertion language for Separation Logic is the logic of bunched implications~\citep{PymOY04}. We use the following grammar:
\begin{equation*}
	\Asl \ni P, Q ::= P \implies Q \mid \exists x . P \mid \code{b} \mid \fwsem{\regr} P \mid \bwsem{\regr} P \mid \emp \mid x \mapsto \code{a} \mid x \dealloc \mid P \andsep Q
\end{equation*}

The interpretation of spacial constructs ($\emp$, $\mapsto$, $\dealloc{}$ and $\andsep$) is as follows. $\emp$ is valid on any state $(s, [])$, independently of the store $s$. $x \mapsto v$ is valid on any state $(s, [s(x) \mapsto v])$, where the heaplet contains only location $s(x)$: this means the memory address stored in variable $x$ points to the value $v$. Similarly, $x \dealloc{}$ holds on $(s, [s(x) \mapsto \delta])$. Finally, the separation conjunction $P \andsep Q$ holds on any state where the heaplet can be split in two sub-heaplets with \emph{disjoint} domains, one satisfying $P$ and the other satisfying $Q$. The disjointness condition ensures that only one of the two sub-formulae can take \emph{ownership} of each location, and it's the key ingredient to enable distinguishing features of separation logics (eg. the Frame rule).

\begin{example}
	Consider the assertion $(\true \andsep v \mapsto - \andsep y \mapsto - \andsep (x = y \lor x \dealloc{}))$ from Example~\ref{ex:intro-push-back}. Using distribution laws of $\andsep$ and $\lor$ we can rewrite it as
	\begin{align*}
		(\true \andsep v \mapsto - \andsep y \mapsto - \andsep x = y) \lor (\true \andsep v \mapsto - \andsep y \mapsto - \andsep x \dealloc{})
	\end{align*}
	On the one hand, the first disjunct explicitly says that $x$ and $y$ are aliased. On the other hand, the other disjunct implicitly says that they are \emph{not} aliased: the separate conjunction $(y \mapsto - \andsep x \dealloc{})$ ensures that the addresses stored in $x$ and $y$ are different. In fact, if they were the same location $l$, it would be impossible to split the heaplet in such a way that $l$ is in the domains of both the (disjoint) sub-heaplets satisfying $y \mapsto -$ and $x \dealloc{}$, respectively.
\end{example}

\begin{figure}[t]
	\centering
	\begin{framed}
		\small
			\(
			\begin{array}{c}
				\infer[\ilrule{Load (ISL)}]
				{\vdash \iltripleo{\oktext{ok: x = x' \andsep y \mapsto e}}{\code{x := [y]}}{x = e[x'/x] \andsep y \mapsto e[x'/x]}}
				{}
				\\[7.5pt]
				\infer[\silrule{Load (SepSIL)}]
				{\vdash \siltriple{\oktext{ok: y \mapsto a \andsep q[a/x]}}{\code{x := [y]}}{\oktext{ok: y \mapsto a \andsep q}}}
				{x \notin \text{fv}(a)}
				\\[7.5pt]

				\infer[\ilrule{StoreEr (ISL)}]
				{\vdash \iltriplee{\oktext{ok: x \dealloc}}{\code{[x] := y}}{x \dealloc{}}}
				{}
				\\[7.5pt]
				\infer[\silrule{StoreEr (SepSIL)}]
				{\vdash \siltriple{\oktext{ok: x \dealloc}}{\code{[x] := y}}{\ertext{er: x \dealloc{}}}}
				{}
				
			\end{array}
			\)
	\end{framed}
	\Description{See caption.}
	\caption{ISL~\citep{RaadBDDOV20} and Separation SIL~\citep{AscariBGL25} rules (excerpt)}\label{fig:sep-il-sil-rules}
\end{figure}

Both IL and SIL have been extended to a separation counterpart, Incorrectness Separation Logic and Separation SIL respectively. They validate the same axioms as their non-separation counterparts, together with rules for the new atomic commands (an excerpt is in Fig.~\ref{fig:sep-il-sil-rules}) and the frame rule:
\[
\infer[\ilrule{Frame (ISL)}]
{\vdash \iltriple{p \andsep f}{\regc}{q \andsep f}}
{\vdash \iltriple{p}{\regc}{q} & \text{comp}(\regc, f)}
\hspace{3em}
\infer[\silrule{Frame (SepSIL)}]
{\vdash \siltriple{p \andsep f}{\regc}{q \andsep f}}
{\vdash \siltriple{p}{\regc}{q} & \text{comp}(\regc, f)}
\]
where $\text{comp}(\regc, f)$ means that command $\regc$ does not modify any of the free variables of assertion $f$.

\subsection{UNTer}\label{sec:unter-bg}
UNTer~\citep{RaadVO24} is a proof system inspired by IL aimed at proving the presence of (non)termination bug. To do so, alongside forward under-approximation (IL) triples, it introduces backwards under-approximation (SIL) triples.
One of the key observation is that forward and backward under-approximation share most of the structural rules, with the most notable exception being the rule of consequence. This leads to a simple automation of backward under-approximation, since it can just reuse most of the reasoning engine already implemented for forward under-approximation (via the use of indexed disjunctions and matched dropping, as described in \citet[\S~2, \textit{Forward versus Backward Under-Approximate Triples}]{RaadVO24}). Particularly, by presenting a \emph{kernel} set of IL-inspired rules \citep[Fig.~1, $\vdash_{\dagger}$ proof system]{RaadVO24} that does not include the rule of consequence, UNTer details a proof system that can prove triples valid for both IL and SIL at the same time.
Such a kernel set contains the same rules as the IL proof system in Fig.~\ref{fig:il-rules} with the following differences: the rule \ilrule{cons} is replaced with dropping of indexed disjuncts and the rule \ilrule{assume} is replaced by the rule
\[
\infer[\text{assume}]{\vdash_{\dagger} \iltriple{P \land b}{\code{b?}}{P \land B}}{}
\]

Similarly, the separation logic instance of UNTer uses the same axioms as ISL, except for the rule of consequence and \ilrule{assume}, changed as above.

UNTer proof system is proved sound (with respect to both validity as IL and SIL triples when excluding the opposite consequence rule) \citep[Th.~7]{RaadVO24}. Moreover, when the consequence rule of IL (resp. SIL) is added to the kernel set of rules, the resulting proof systems becomes also complete for IL (resp. SIL) \citep[Th.~8]{RaadVO24}. To our knowledge, there is no completeness result concerning only the kernel set of rules with respect to triples that are valid both for IL and SIL at the same time.

\section{Forward/backward axioms for atomic commands}\label{sec:atoms}
In UNTer~\citep[Fig.~7]{RaadVO24} the authors proves that IL (and ISL) axioms for atomic commands are also valid as SIL triples. However, these axioms are based on Incorrectness Separation Logic and hand-crafted. Therefore, it is natural to ask the following two questions.
\begin{enumerate}
	\item Are these axiom as general as possible?
	\item Is there a general procedure to derive axioms for new atomic commands, not relying on pre-existing ISL axioms?
\end{enumerate}

The first question is partially answered in the positive by UNTer completeness result~\citep[\S~6]{RaadVO24}: since the resulting proof system is complete, each axiom together with the rule of consequence of IL (resp. SIL) is able to prove every valid IL (resp. SIL) triple for that particular atomic command.
However, nothing is said about completeness with respect to triples that are both IL and SIL at the same time: is it possible to prove any such triple without resorting to the consequence rule of either logic (which can make the triple unsound for the other logic)?

We tackle this problem by addressing the second question. More in details, we propose an axiom \emph{schema} for atomic commands that is sound and complete for triples that are both IL and SIL. From this, we derive axioms for atomic commands that are sound and complete by construction, and we then show that we can derive such axioms in UNTer.
However, our contribution goes beyond this: by providing this general schema, we give a methodology that can be applied to any atomic command. As an example, we will consider non-deterministic assignment, a command missing in UNTer, and derive a new axiom for it.

\begin{proposition}\label{prop:axiom-schema-valid}
	For every command $\regc$ and assertions $P$, $Q$, the pre $(P \land \bwsem{\regc} Q)$ and the post $(Q \land \fwsem{\regc} P)$ makes both a valid IL and SIL triple:
	\[
	\vDash \iltriple{P \land \bwsem{\regc} Q}{\regc}{Q \land \fwsem{\regc} P} \qquad\land\qquad \vDash \siltriple{P \land \bwsem{\regc} Q}{\regc}{Q \land \fwsem{\regc} P}
	\]
\end{proposition}

We show below some examples of applications of this schema. Note that we will often replace the (forward or backward) semantics with the equivalent formula from Section~\ref{sec:assertion-language}.
\begin{example}
	For assignments, the above schema yields the axiom:
	\[
	\iltriple{P \land Q[a / x]}{\code{x := a}}{Q \land \spa{x}{a}{P}}\ .
	\]
	This is equivalent to the UNTer/IL axiom, that is precisely Floyd's forward axiom.
	Setting $Q = \true$ in our axiom yields precisely the UNTer axiom.
	Conversely, substituting $P$ with $(P \land Q[a / x])$ in the UNTer axiom yields ours, after some equivalence-preserving transformation of the formula in the post.
	
	For Boolean guards, the above schema yields the axiom:
	\[
	\iltriple{P \land Q \land b}{\code{b?}}{P \land Q \land b}\ .
	\]
	Again, to derive the UNTer axiom $\vdash_{\dagger} \iltriple{P \land b}{\code{b?}}{P \land b}$ it suffices to take $Q = \true$ in our axiom.
	Conversely, substituting $P$ with $(P \land Q)$ in the UNTer axiom yields ours.
\end{example}

To show how our schema can be used to handle new constructs, we consider nondeterministic assignments, which was not explicitly discussed in UNTer.
\begin{example}
	Recalling that for nondeterministic assignment
	\[
	\fwsem{\code{x := nondet()}} P = \exists x . P \qquad\qquad \bwsem{\code{x := nondet()}} Q = \exists x . Q
	\]
	we obtain the axiom
	\[
	\iltriple{P \land \exists x . Q}{\code{x := nondet()}}{Q \land \exists x . P}
	\]
	that is new and stronger than other proposals. In UNTer, nondeterministic assignments are not present in the programming language.
	IL and SIL use, respectively, the axioms
	\[
	\vdash \iltriple{P}{\code{x := nondet()}}{\exists x . P} \qquad\qquad \vdash \siltriple{\exists x . Q}{\code{x := nondet()}}{Q}
	\]
	which can be combined in
	\[
	\iltriple{\exists x . P}{\code{x := nondet()}}{\exists x . P}
	\]
	However, this latter axiom is weaker than the proposal obtained with our methodology. For instance, it cannot be exploited to prove the triple $\iltriple{\true}{\code{x := nondet()}}{x > 0}$.
	
	Interestingly, ISL~\citep{RaadBDDOV20} uses the axiom \ilrule{havoc}
	\[
	\vdash \iltriple{x = n}{\code{x := nondet()}}{x = m}
	\]
	that is equivalent to ours. In fact, assuming $n$ and $m$ are free names in $P$, $Q$, we can prove our axiom from \ilrule{havoc} (together with \ilrule{frame} and \ilrule{exist}) with the following derivation:
	\[
	\infer[\ilrule{exist}]{\iltriple{P \land \exists m . Q[m / x]}{\code{x := nondet()}}{Q \land \exists n . P[n / x]}}{
	\infer[\ilrule{frame}]{\iltriple{P \land Q[m / x]}{\code{x := nondet()}}{x = m \land \exists n . P[n / x] \land Q[m / x]}}{
	\infer[\ilrule{exist}]{\iltriple{\exists n . (x = n \land P[n / x])}{\code{x := nondet()}}{\exists n . (x = m \land P[n / x])}}{
	\infer[\ilrule{frame}]{\iltriple{x = n \land P[n / x]}{\code{x := nondet()}}{x = m \land P[n / x]}}{
	\infer[\ilrule{havoc}]{\iltriple{x = n}{\code{x := nondet()}}{x = m}}{}	
	}	}	}	}
	\]
	Conversely, we can derive \ilrule{havoc} from our axiom just by taking $P = (x = n)$ and $Q = (x = m)$.
	More abstractly, we know the valid triple $\iltriple{x = n}{\code{x := nondet()}}{x = m}$ (as well as any other valid triple) is derivable by completeness of our axiom schema (Proposition~\ref{prop:axiom-schema-complete} below).
\end{example}
\begin{example}
	Let us consider an artificial example to show how we can apply our schema to new expressions. Consider a new atomic command $\code{x++?}$ that nondeterministically can opt to increment \code{x} or leave it unchanged. Semantically, it is equivalent to the code \code{skip $\regplus$ (x := x+1)}. From this, we derive its forward and backward semantics
	\[
	\fwsem{\code{x++?}} P = P \lor P[x - 1 / x]
	\qquad\qquad
	\bwsem{\code{x++?}} Q = Q \lor Q[x + 1 / x]
	\]
	so that our schema readily yields the axiom
	\[
	\iltriple{P \land (Q \lor Q[x + 1 / x])}{\code{x++?}}{Q \land (P \lor P[x - 1 / x])}
	\]
\end{example}

This axiom schema is also complete. To prove this, we show that any triple valid for both IL and SIL can be rewritten as $\iltriple{P \land \bwsem{\regc} Q}{\regc}{Q \land \fwsem{\regc} P}$, thus being provable with our axiom schema.
\begin{proposition}\label{prop:axiom-schema-complete}
	For every command $\regc$ and sets of states $P$, $Q$, if both the IL triple $\vDash \iltriple{P}{\regc}{Q}$ and the SIL triple $\vDash \siltriple{P}{\regc}{Q}$ are valid, then both $P \land \bwsem{\regc} Q = P$ and $Q \land \fwsem{\regc} P = Q$.
\end{proposition}

\subsection{Heap manipulating axioms}
The result in the previous section considers a simple, imperative language. In theory, the approach can be extended directly to heap-manipulating commands by changing the semantics. However, this approach does not take into account the locality principle of separation logic, according to which one should define \emph{small} axioms---whose pre- and postconditions deals with the minimal amount of information needed to execute the command---that can be extended by need to larger  heaps thanks to a suitable \emph{frame} rule, the hallmark of separation logics.

To recover local axioms, we can consider a local semantics $\fwsemL{\cdot}$ instead of the global $\fwsem{\cdot}$. We define such a semantics based on the relation \textbf{foot} in \citet[\S~4.1]{RaadBDDOV20}. Intuitively, $\textbf{foot}(\regc)$ relates a pair of states $(s, h)$, $(s', h')$ if executing $\regc$ starting from $(s, h)$ can yield the final state $(s', h')$ and $h$ is a minimal heaplet allowing for such an execution. In other words, if we remove any location from $h$ then the command $\regc$ can no longer execute from the reduced state.

We can then define the local semantics $\fwsemL{\regc}$ as the functional version of the \textbf{foot} relation: $\fwsemL{\regc} (s, h) = \{ (s', h') \mid ((s, h), (s', h')) \in \textbf{foot}(\regc) \}$ and then extended by additivity to sets of states. Leveraging their footprint theorem \citep[Th.~2]{RaadBDDOV20}, we obtain an analogous decomposition of the (global) semantics in terms of local semantics and frames:
\begin{proposition}
	For any command $\regc$, assertions $P$, $R$ such that $\fwsemL{\regc} \sigma$ is defined for every $\sigma \in P$
	\[
	\fwsem{\regc} (P \andsep R) = (\fwsemL{\regc} P) \andsep R
	\]
\end{proposition}

From this, we obtain a ``local axiom schema'' for heap manipulating commands:
\begin{proposition}
	For every command $\regc$ and assertions $P$, $Q$, the pre $(P \land \bwsemL{\regc} Q)$ and the post $(Q \land \fwsemL{\regc} P)$ makes both a valid ISL and Separation SIL triple:
	\[
	\vDash \iltriple{P \land \bwsemL{\regc} Q}{\regc}{Q \land \fwsemL{\regc} P} \qquad\land\qquad \vDash \siltriple{P \land \bwsemL{\regc} Q}{\regc}{Q \land \fwsemL{\regc} P}
	\]
\end{proposition}
The proof is identical to that of Proposition~\ref{prop:axiom-schema-valid} by recalling that $\bwsemL{\regc}$ is a subset of $\bwsem{\regc}$. Moreover, the locality of $\bwsemL{\cdot}$ forces locality in the axiom thanks to the conjunction $\land$: even if, for instance, $P$ talks about locations outside the footprint of $\regc$, these are filtered out by the $\bwsemL{\regc} Q$ conjunct in the precondition, forcing locality.

As an example, we apply our schema to derive the axiom for a load command.
\begin{example}
	Consider a load command $\code{x := [y]}$. Its local forward and backward semantics are:
	\begin{align*}
		\fwsemL{\code{x := [y]}} (y \mapsto v \land P) &= y \mapsto v \land \exists z . (P[z / x] \land x = v) \\
		\bwsemL{\code{x := [y]}} (y \mapsto v \land Q) &= y \mapsto v \land Q[v / x]
	\end{align*}
	Note the additional conjunct $y \mapsto v$ in the input states: this ensures that the the heap(let) is only defined on the location pointed by $y$, that is exactly the footprint of the load statement.
	
	Our schema applied to these semantics yields the triple
	\[
	\iltriple{P \land (y \mapsto v \land Q[v / x]))}{\code{x := [y]}}{Q \land (y \mapsto v \land \exists z . (P[z / x] \land x = v))}
	\]
	which can be simplified to
	\[
	\iltriple{y \mapsto v \land P \land Q[v / x]}{\code{x := [y]}}{y \mapsto v \land x = v \land Q \land \exists z . P[z / x]}
	\]
	From this, we can recover the load axiom in ISL/UNTer~\citep[Fig.~10]{RaadVO24extended} (both use the same axiom). To do so, we first instance our axiom by taking $Q \eqdef (\true)$ and $P \eqdef (x = x' \land v = e)$, and then use rule $\ilrule{exists}$ to hide $v$.
	The precondition simplifies as
	\begin{align*}
		& \exists v . (y \mapsto v \land (x = x' \land v = e) \land (\true)[v / x]) \\
		\equiv\; & \exists v . (y \mapsto e \land x = x' \land v = e) \\
		\equiv\; & y \mapsto e \land x = x'
	\end{align*}
	and the postcondition simplifies as
	\begin{align*}
		& \exists v . (y \mapsto v \land x = v \land (\true) \land \exists z . (x = x' \land v = e)[z / x]) \\
		\equiv\; & \exists v . (y \mapsto v \land x = v \land \exists z . (z = x' \land v = e[z / x])) \\
		\equiv\; & \exists v . (y \mapsto v \land x = v \land v = e[x' / x]) \\
		\equiv\; & y \mapsto e[x' / x] \land x = e[x' / x]
	\end{align*}
	obtaining exactly the UNTer axiom.

\end{example}

\section{U-turn: Following IL Derivations with SIL}\label{sec:uturn}

The approach in the previous section provides a technique to derive axioms that are valid in both IL and SIL. Paired with UNTer, it enables a single analysis resulting in a triple with a double guarantee: all errors in the post are reachable from states in the pre and all states in the pre can lead to some error in the post. However, our axiom schema requires previous knowledge of \emph{both} $P$ and $Q$, that is both the pre and the post of the expected triple or at least some over-approximation of them. Since the analysis typically follows the control flow either in the forward or backward direction, it is often the case that only one of the two is available (the pre in a forward analysis and the post in a backward one). A possible solution would be to use a default value (such as $\true$) for the unknown pre- or postcondition.

In this section, we tackle the problem from a completely different angle. Instead of doing a single analysis, whose result is valid both for IL and SIL but that is tied to either the forward or backward flow in its computation, we perform two consecutive analyses. We start with a forward, IL-based analysis, and then we trace it backwards using SIL principles. In doing so, we take advantage of the information discovered during the forward analysis. We call \textbf{U-Turn} the resulting proof system.

Intuitively, each IL derivation outlines those \emph{code paths} that have been explored to find the result. For instance, if the proof uses $\ilrule{choiceL}$ to analyse an \code{if} statement, it means that we are only considering the then-branch path in the code, dropping the analysis of the else-branch. Similarly, usage of rules \ilrule{iter0} and \ilrule{unroll} details how many loop unrolling have been performed. This information is incredibly valuable for a backward step with SIL, because it guides the proof search down paths that are \emph{guaranteed to succeed}. While this strategy does not ensure completeness by itself (i.e., we may still miss some sources of the errors in the post) it is often useful to report \emph{some} sufficient preconditions for the errors rather than aiming to collect all of them.

To formally develop this idea, we consider U-Turn judgments of the form

\[
\judgprq
\]

\noindent
where $d$ is a \emph{proof tree}, built from the rules in Fig.~\ref{fig:il-rules}, for the provable IL triple $\vdash \iltriple{P}{\regr}{Q}$. As discussed above the derivation $d$ contains useful information on the code paths that is not summarized in the final IL triple. We will discuss how such reasoning is implemented by some U-turn rules.

\begin{definition}[Judgment validity]\label{def:uturn-validity}
	Given a proof tree $d$ for the provable IL triple $\vdash \iltriple{P}{\regr}{Q}$ (using the IL proof system in Fig.~\ref{fig:il-rules}), we say that the U-Turn judgment $\ \judgprq$ is \emph{valid}, written $\judgvprq$,
	if 
	\begin{enumerate}
		\item $\vDash \siltriple{P'}{\regr}{Q'}$,
		\item $P' \subseteq P$,
		\item $Q' \subseteq Q$,
		\item either $P' = Q' = \emptyset$ or both $P', Q' \neq \emptyset$.
	\end{enumerate}
\end{definition}

A valid U-Turn judgment entails the validity of the corresponding SIL triple $\siltriple{P'}{\regr}{Q'}$ (condition~(1)) and that both $P'$ and $Q'$ are subsets of the corresponding IL pre/posts (conditions~(2) and~(3)). In the post, this inclusion means we are allowed to only focus on a subset $Q'$ of the states found in the post $Q$. This freedom is mostly a technical requirement to be used inside derivations rather than to drop errors found by the IL analysis, for instance to drop some non-interesting \oktext{ok} states.
This requirement is immediately evident when considering, e.g., the analysis of two consecutive code fragments: given the IL derivation:
\[
\ilderiv{\ilderiv{d_1}{P}{\regr_1}{R} & \ilderiv{d_2}{R}{\regr_2}{Q}}{P}{\regr_1; \regr_2}{Q}
\]
\noindent
we use SIL to trace back the sources of errors in $Q$ that reside in $R$ w.r.t. executing $\regr_2$, which may lead to a proper subset $R'\subset R$ of IL postcondition for $\regr_1$ in 
\[
\ilderiv{d_1}{P}{\regr_1}{R}\ .
\]
\noindent
Whence the need to, inductively, being able to start the inference process in SIL along $\regr_1$ starting from any subset $R'$ of $R$ rather than from $R$ itself.
Condition (4) is a bit more involved. If $Q'$ is not empty we care about reachability of some final state in $Q$, and forcing $P'$ to be non-empty means we find (some) states in $P$ that surely lead to errors in $Q$. If instead $Q'$ is empty it means we are not considering any of the states found by the IL analysis in the post $Q$, so by taking $P' = \emptyset$ we ignore completely the program path that ends at $Q'$. This gives us the freedom to drop some of the code paths explored in the IL triple if we deem them not interesting.
Formally, condition (4) is justified by Lemma~\ref{lmm:p-q-valid-empty}.1: if $Q'$ is empty, $P'$ must be empty as well since we require $\vDash \siltriple{P'}{\regr}{Q'}$.

\begin{figure}[t]
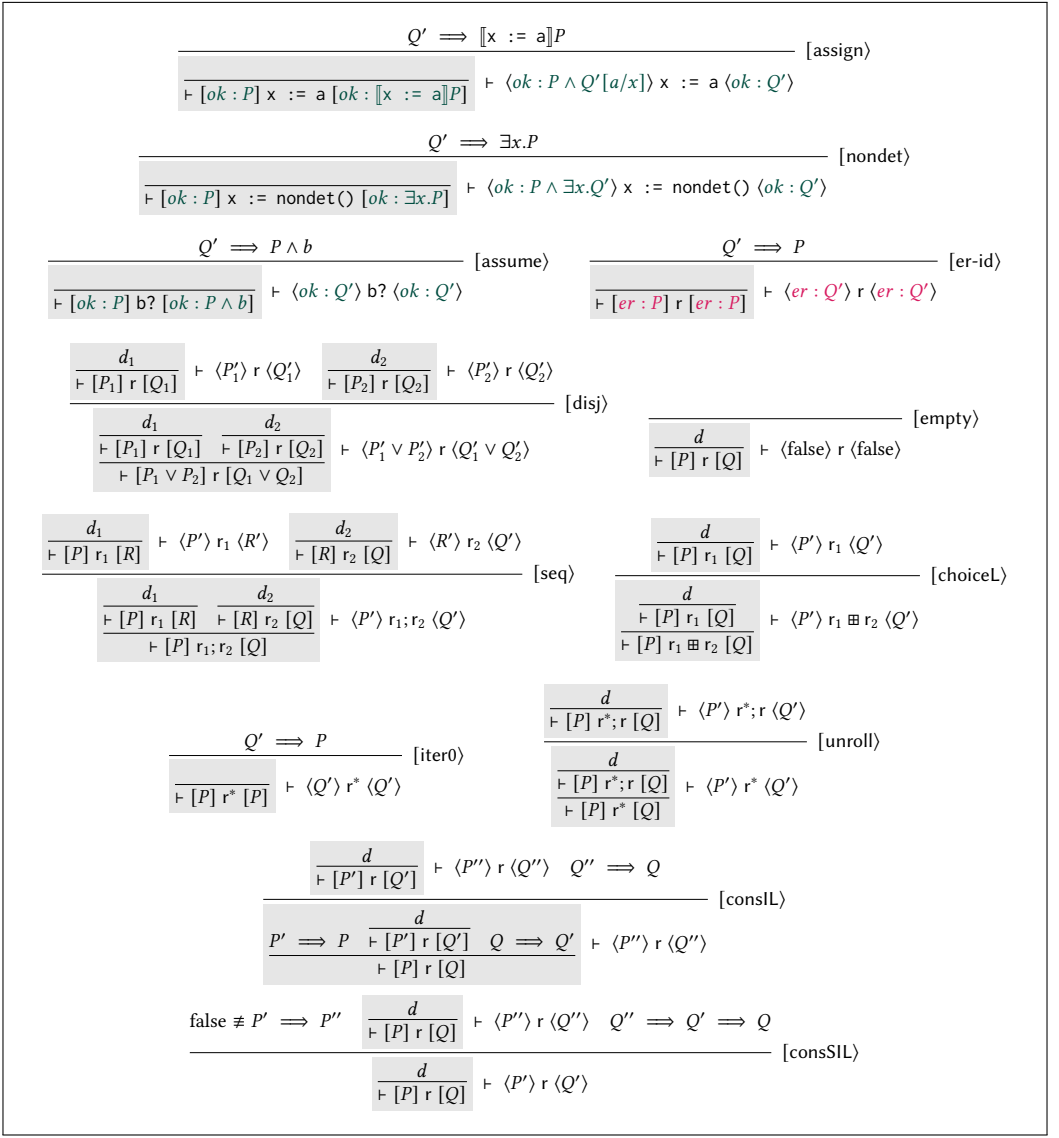

	\centering
	\begin{framed}
	\resizebox{\textwidth}{!}{
			\(
			\begin{array}{c}
				\infer[\lrule{assign}]
				{ \ilsiljudgment{\ilderiv{\phantom{d}}{\oktext{ok: P}}{\code{x := a}}{\oktext{ok: \spa{x}{a}{P}}}}{\oktext{ok: P \land Q'[a / x]}}{\code{x := a}}{\oktext{ok: Q'}}}
				{ Q' \implies \spa{x}{a}{P} }
				\\[4.5ex]
				\infer[\lrule{nondet}]
				{ \ilsiljudgment{\ilderiv{\phantom{d}}{\oktext{ok: P}}{\code{x := nondet()}}{\oktext{ok: \exists x . P}}}{\oktext{ok: P \land \exists x . Q'}}{\code{x := nondet()}}{\oktext{ok: Q'}} }
				{ Q' \implies \exists x . P }
				\\[4.5ex]
				\infer[\lrule{assume}]
				{ \ilsiljudgment{\ilderiv{\phantom{d}}{\oktext{ok: P}}{\code{b?}}{\oktext{ok: P \land b}}}{\oktext{ok: Q'}}{\code{b?}}{\oktext{ok: Q'}} }
				{ Q' \implies P \land b }
				\qquad
				\infer[\lrule{er\mbox{-}id}]
				{ \ilsiljudgment{ \ilderiv{\phantom{d}}{\ertext{er: P}}{\regr}{\ertext{er: P}} }{\ertext{er: Q'}}{\regr}{\ertext{er: Q'}} }
				{ Q' \implies P }
				\\[4.5ex]
				\infer[\lrule{disj}]
				{ \ilsiljudgment{\ilderiv{\ilderiv{d_1}{P_1}{\regr}{Q_1} & \ilderiv{d_2}{P_2}{\regr}{Q_2}}{P_1 \lor P_2}{\regr}{Q_1 \lor Q_2}}{P_1' \lor P_2'}{\regr}{Q_1' \lor Q_2'} }
				{ \ilsiljudgment{\ilderiv{d_1}{P_1}{\regr}{Q_1}}{P_1'}{\regr}{Q_1'} & \ilsiljudgment{\ilderiv{d_2}{P_2}{\regr}{Q_2}}{P_2'}{\regr}{Q_2'} }
				\qquad
				\infer[\lrule{empty}]
				{ \ilsiljudgment{\ilderiv{d}{P}{\regr}{Q}}{\false}{\regr}{\false} }
				{ }
				\\[6ex]
				\infer[\lrule{seq}]
				{ \ilsiljudgment{\ilderiv{\ilderiv{d_1}{P}{\regr_1}{R} & \ilderiv{d_2}{R}{\regr_2}{Q}}{P}{\regr_1; \regr_2}{Q}}{P'}{\regr_1; \regr_2}{Q'} }
				{ \ilsiljudgment{\ilderiv{d_1}{P}{\regr_1}{R}}{P'}{\regr_1}{R'} & \ilsiljudgment{\ilderiv{d_2}{R}{\regr_2}{Q}}{R'}{\regr_2}{Q'} }
				\qquad
				\infer[\lrule{choiceL}]
				{ \ilsiljudgment{\ilderiv{\ilderiv{d}{P}{\regr_1}{Q}}{P}{\regr_1 \regplus \regr_2}{Q}}{P'}{\regr_1 \regplus \regr_2}{Q'} }
				{ \ilsiljudgment{\ilderiv{d}{P}{\regr_1}{Q}}{P'}{\regr_1}{Q'} }
				\\[6ex]
				\infer[\lrule{iter0}]
				{ \ilsiljudgment{\ilderiv{\phantom{d}}{P}{\regr^{\kstar}}{P}}{Q'}{\regr^{\kstar}}{Q'} }
				{ Q' \implies P }
				\qquad\qquad %
				\infer[\lrule{unroll}]
				{ \ilsiljudgment{\ilderiv{\ilderiv{d}{P}{\regr^{\kstar}; \regr}{Q}}{P}{\regr^{\kstar}}{Q}}{P'}{\regr^{\kstar}}{Q'} }
				{ \ilsiljudgment{\ilderiv{d}{P}{\regr^{\kstar}; \regr}{Q}}{P'}{\regr^{\kstar}; \regr}{Q'} }
				\\[5ex]
				\infer[\lrule{consIL}]
				{ \ilsiljudgment{\ilderiv{P' \implies P & \ilderiv{d}{P'}{\regr}{Q'} & Q \implies Q'}{P}{\regr}{Q}}{P''}{\regr}{Q''} }
				{ \ilsiljudgment{\ilderiv{d}{P'}{\regr}{Q'}}{P''}{\regr}{Q''} & Q'' \implies Q }
				\\[4.5ex]
				\infer[\lrule{consSIL}]
				{ \ilsiljudgment{\ilderiv{d}{P}{\regr}{Q}}{P'}{\regr}{Q'} }
				{ \false \nequiv P' \implies P'' & \ilsiljudgment{\ilderiv{d}{P}{\regr}{Q}}{P''}{\regr}{Q''} & Q'' \implies Q' \implies Q }
			\end{array}
			\)
		}
	\end{framed}
	\Description{See caption.}
	\caption{The U-Turn proof system (excerpt). The full proof system can be found in the Appendix, Fig.~\ref{fig:app:uturn-rules-full}.}\label{fig:uturn-rules}
\end{figure}

As a main contribution, we define the U-Turn proof system for such judgments. A relevant excerpt on which we focus is given in Fig.~\ref{fig:uturn-rules}. The full proof system is available in the Appendix, Fig.~\ref{fig:app:uturn-rules-full}. Most U-Turn rules can only be applied when the IL derivation $d$ ends with the application of a specific IL rule, and in that case they share its name. This constraint means that the SIL derivation in the U-Turn proof system will mimic the IL one.

Rule \lrule{assign} can only be applied when IL uses its own axiom \ilrule{assign}. It allows one to take any subset of the strongest post and go backward from there, then conjoin it with $P$ to ensure the $P' \subseteq P$ validity condition of the triple. Rules \lrule{nondet} and \lrule{assume} work similarly, and analogous rules are available for other atoms.

If the IL derivation exploited \ilrule{disj} to split the precondition in two disjuncts and analyse them separately, U-Turn requires the SIL step to do the same: it analyses the two resulting post $Q_1$ and $Q_2$ separately and then joins the results.
Similarly, whenever the IL derivation composes sequentially two sub-proofs using \ilrule{seq}, U-Turn forces SIL to do the same with the rule \lrule{seq}.
Moreover, if IL found that the error originated before $\regr$ and just propagated it through the command with \ilrule{er\mbox{-}id}, the corresponding U-Turn rule \lrule{er\mbox{-}id}  propagates backward the error as-is.

Rule \lrule{empty} is peculiar in that it can be applied regardless of the IL derivation. However, it can only be applied when the post is $\false$,  and it derives the (only) valid pre $\false$. Intuitively, this correspond to not analysing $\regr$ when we don't care about any state it can reach.

Rules \lrule{choiceL}, \lrule{iter0} and \lrule{unroll} show how U-Turn force SIL to follow the same code paths analysed by IL. In all three rules, if IL decides to follow a specific code path (the left branch in an if, skip a loop or unroll it once) then SIL is forced to follow the exact same path.

If the IL derivation contains any application of its rule of consequence \ilrule{cons}, U-Turn skips it with \lrule{consIL}: when using the proof system for a backward analysis, the constraint $Q'' \implies Q$ will always be satisfied because $Q''$ will be the pre in some subsequent code fragment whose pre in the IL triple is $Q$. Intuitively, since IL rule of consequence does not change the explored code paths, it is not relevant for U-Turn.

Lastly, at any point in the derivation, U-Turn can use SIL rule of consequence via \lrule{consSIL}, provided it does not weakens the triple so much that it breaks one of the validity conditions of U-Turn.
As we will discuss later, if rule \lrule{consSIL} is never used, then we will have stronger guarantees about the result of the analysis (see Theorem~\ref{th:uturn-il-valid}).

The U-Turn proof system is sound. The proof is a standard induction on the derivation tree.
\begin{theorem}[Soundness]\label{th:uturn-soundness}
	Any provable U-Turn judgment is valid.
\end{theorem}

We present now two examples of how U-Turn guides the SIL proof search.

\begin{example}\label{ex:uturn-string-null}
\begin{figure}[t]
\begin{minipage}[t]{0.25\textwidth}
\begin{lstlisting}[language=C]
// program rlen
s := init(l);
l := 100;
i := len(s)
\end{lstlisting}
\end{minipage}
\begin{minipage}[t]{0.33\textwidth}
\begin{lstlisting}[language=C]
init(l) {
  s := alloc(l + 1);
  // initialize s[0..l-1]
  if (l != 3) {
    s[l] := "\0";
  }
  return s
}
\end{lstlisting}
\end{minipage}
\begin{minipage}[t]{0.35\textwidth}
\begin{lstlisting}[language=C]
len(s) {
  i := 0;
  while (s[i] != "\0") {
    i := i + 1;
  }
  return i
}
\end{lstlisting}
\end{minipage}
\caption{The program $\mathsf{rlen}$ discussed in Example~\ref{ex:uturn-string-null}.}
\label{fig:rlen}
\end{figure}

	\begin{figure}[t]
		\begin{subfigure}[T]{0.47\textwidth}
			\scriptsize
			\begin{align*}
				&\ilexact{\oktext{ok: \exists \vec{a} . s \mapsto \vec{a} \land l = 100 \land pv(2) \land i = 0 \land si = a0}} \\
				&\quad \code{(si != \nulltc)?;} \\
				&\ilexact{\oktext{ok: \exists \vec{a} . s \mapsto \vec{a} \land l = 100 \land pv(2) \land i = 0 \land si = a0}} \\
				&\quad \code{i := i + 1;} \\
				&\ilexact{\oktext{ok: \exists \vec{a} . s \mapsto \vec{a} \land l = 100 \land pv(2) \land i = 1 \land si = a0}} \\
				&\quad \code{si := s[i]} \\
				&\ilexact{\oktext{ok: \exists \vec{a} . s \mapsto \vec{a} \land l = 100 \land pv(2) \land i = 1 \land si = a1}}
			\end{align*}
			\caption{Linearized ISL proof for the first iteration of the body of the while loop in \code{len}.}
		\end{subfigure}
		\hfill
		\begin{subfigure}[T]{0.47\textwidth}
			\scriptsize
			\begin{align*}
				&\ilexact{\oktext{ok: \exists \vec{a} . s \mapsto \vec{a} \land l = 100 \land pv(2) \land i = 3 \land si = a3}} \\
				&\quad \code{(si != \nulltc)?;} \\
				&\ilexact{\oktext{ok: \exists \vec{a} . s \mapsto \vec{a} \land l = 100 \land pv(3) \land i = 3 \land si = a3}} \\
				&\quad \code{i := i + 1;} \\
				&\ilexact{\oktext{ok: \exists \vec{a} . s \mapsto \vec{a} \land l = 100 \land pv(3) \land i = 4 \land si = a3}} \\
				&\quad \code{si := s[i]} \\
				&\ilexact{\ertext{er: \exists \vec{a} . s \mapsto \vec{a} \land l = 100 \land pv(3) \land i = 4 \land si = a3}}
			\end{align*}
			\caption{Linearized ISL proof for the last iteration of the body of the while loop in \code{len}.}
		\end{subfigure}
		\begin{subfigure}[T]{\textwidth}
			\scriptsize
			\begin{align*}
				&\ilexact{\oktext{ok: \exists \vec{a} . s \mapsto \vec{a} \land l = 100 \land pv(2)}} \\
				&\quad \code{i := 0;} \\
				&\ilexact{\oktext{ok: \exists \vec{a} . s \mapsto \vec{a} \land l = 100 \land pv(2) \land i = 0}} \\
				&\quad \code{si := s[i];} \\
				&\ilexact{\oktext{ok: \exists \vec{a} . s \mapsto \vec{a} \land l = 100 \land pv(2) \land i = 0 \land si = a0}} \\
				&\quad \regr_b; \\
				&\ilexact{\oktext{ok: \exists \vec{a} . s \mapsto \vec{a} \land l = 100 \land pv(2) \land i = 1 \land si = a1}} \\
				&\quad \regr_b; \\
				&\ilexact{\oktext{ok: \exists \vec{a} . s \mapsto \vec{a} \land l = 100 \land pv(2) \land i = 2 \land si = a2}} \\
				&\quad \regr_b; \\
				&\ilexact{\oktext{ok: \exists \vec{a} . s \mapsto \vec{a} \land l = 100 \land pv(2) \land i = 3 \land si = a3}} \\
				&\quad \regr_b; \\				
				&\ilexact{\ertext{er: \exists \vec{a} . s \mapsto \vec{a} \land l = 100 \land pv(3) \land i = 4 \land si = a3}} \\
				&\quad \code{(si == \nulltc)?} \\
				&\ilexact{\ertext{er: \exists \vec{a} . s \mapsto \vec{a} \land l = 100 \land pv(3) \land i = 4 \land si = a3}}
			\end{align*}
			\caption{Sketch of the ISL proof for \code{len}. We call $\regr_b$ the body of the while loop. Since by using \ilrule{unroll} and \ilrule{iter0} the proof unrolls the loop 4 times, we do the same here: hence, the four repetitions of $\regr_b$ instead of $\regr_b^{\kstar}$.}
		\end{subfigure}
		\begin{subfigure}[T]{\textwidth}
			\scriptsize
			\begin{align*}
				&\ilexact{\oktext{ok: \true}} \\
				&\quad \code{s := init(l);} \\
				&\ilexact{\oktext{ok: \exists \vec{a} . s \mapsto \vec{a} \land l = 3 \land pv(2)}} \\
				&\quad \code{l := 100;} \\
				&\ilexact{\oktext{ok: \exists \vec{a} . s \mapsto \vec{a} \land l = 100 \land pv(2)}} \\
				&\quad \code{i := len(s);} \\
				&\ilexact{\ertext{er: \exists \vec{a} . s \mapsto \vec{a} \land l = 100 \land i = 4 \land pv(3)}}
			\end{align*}
			\caption{Sketch of the ISL proof for $\mathsf{rlen}$. We hide the local variable $si$ of \code{len} using ISL rule \ilrule{local}.}
		\end{subfigure}
		\caption{Sketch of the ISL derivation for $\iltriple{\true}{\mathsf{rlen}}{\ertext{er: \exists \vec{a} . s \mapsto \vec{a} \land l = 100 \land i = 4 \land pv(3)}}$.}
		\label{fig:isl-deriv-rlen}
	\end{figure}
	
	For this example we consider the separation logic instance of both IL and SIL (ISL and Separation SIL respectively). Since structural rules are the same as IL and SIL, we only have to adapt atoms, which is straightforward using the corresponding atoms from the two logics and add the frame rule. We also assume both logics include arrays, the extension being straightforward (see, e.g., the treatment in \citet{Reynolds02}).

	Consider the faulty program $\mathsf{rlen}$ in Fig.~\ref{fig:rlen}. First, it initializes a string $s$ with length $l$. However, when $l = 3$, it misses the null terminator. Then, a client tries to compute the length of the string, iterating over it and looking for the null terminator. This makes the bug emerge whenever the initial value of $l$ is $3$, but this information is obfuscated after the assignment \code{l := 100}.

	For presentation purposes, we write $\vec{a}$ for $a0, a1, a2, a3$ and $pv(n)$ for $(a0 \neq \nullt \land \dots \land an \neq \nullt)$. Since our syntax doesn't allow variable dereferencing in boolean expressions, we desugar the guard of the while-loop \code{(s[i] != \nulltc)} using a temporary variable \code{si} that is assigned to \code{si := s[0]} before the loop and to \code{si := s[i]} inside it.
	Using ISL, we can prove the following triple as shown in Fig.~\ref{fig:isl-deriv-rlen}. 
	\[
	\iltriple{\oktext{\okflag: \true}}{\mathsf{rlen}}{\ertext{er: \exists \vec{a} . s \mapsto \vec{a} \land l = 100 \land i = 4 \land pv(3)}}
	\]

	\begin{figure}[t]
		\begin{subfigure}[T]{\textwidth}
			\scriptsize
			\begin{align*}
				\uturnderivassertion{\oktext{ok: \exists \vec{a} . s \mapsto \vec{a} \land l = 100 \land pv(2) \land i = 3 \land si = a3}}
					{\oktext{ok: \exists \vec{a} . s \mapsto \vec{a} \land l = 100 \land pv(3) \land i = 3 \land si = a3 \land si \neq \nullt}} \\
				\uturnderivcode{\code{(si != \nulltc)?;}} \\
				\uturnderivassertion{\oktext{ok: \exists \vec{a} . s \mapsto \vec{a} \land l = 100 \land pv(3) \land i = 3 \land si = a3}}
					{\oktext{ok: \exists \vec{a} . s \mapsto \vec{a} \land l = 100 \land pv(3) \land i = 3 \land si = a3}} \\
				\uturnderivcode{\code{i := i + 1;}} \\
				\uturnderivassertion{\oktext{ok: \exists \vec{a} . s \mapsto \vec{a} \land l = 100 \land pv(3) \land i = 4 \land si = a3}}
					{\oktext{ok: \exists \vec{a} . s \mapsto \vec{a} \land l = 100 \land pv(3) \land i = 4 \land si = a3}} \\
				\uturnderivcode{\code{si := s[i]}} \\
				\uturnderivassertionb{\ertext{er: \exists \vec{a} . s \mapsto \vec{a} \land l = 100 \land pv(3) \land i = 4 \land si = a3}}
					{\ertext{er: \exists \vec{a} . s \mapsto \vec{a} \land l = 100 \land pv(3) \land i = 4 \land si = a3}}
			\end{align*}
			\caption{Linearized U-Turn proof for the last iteration of the body of the while loop in \code{len}. Note that $si \neq \nullt$ in the first line is redundant since $si = a2$ and $pv(3)$ contains $a2 \neq \nullt$.}
			\label{fig:il+sil-deriv-rlen:1}
		\end{subfigure}
		\begin{subfigure}[T]{\textwidth}
			\scriptsize
			\begin{align*}
				\uturnderivassertion{\oktext{ok: \exists \vec{a} . s \mapsto \vec{a} \land l = 100 \land pv(2)}}
					{\oktext{ok: \exists \vec{a} . s \mapsto \vec{a} \land l = 100 \land pv(3) \land 0 = 0}} \\
				\uturnderivcode{\code{i := 0;}} \\
				\uturnderivassertion{\oktext{ok: \exists \vec{a} . s \mapsto \vec{a} \land l = 100 \land pv(2) \land i = 0}}
					{\oktext{ok: \exists \vec{a} . s \mapsto \vec{a} \land l = 100 \land pv(3) \land i = 0 \land a0 = a0}} \\
				\uturnderivcode{\code{si := s[i];}} \\
				\uturnderivassertion{\oktext{ok: \exists \vec{a} . s \mapsto \vec{a} \land l = 100 \land pv(2) \land i = 0 \land si = a0}}
					{\oktext{ok: \exists \vec{a} . s \mapsto \vec{a} \land l = 100 \land pv(3) \land i = 0 \land si = a0}} \\
				\uturnderivcode{\regr_b;} \\
				\uturnderivassertion{\oktext{ok: \exists \vec{a} . s \mapsto \vec{a} \land l = 100 \land pv(2) \land i = 1 \land si = a1}}
					{\oktext{ok: \exists \vec{a} . s \mapsto \vec{a} \land l = 100 \land pv(3) \land i = 1 \land si = a1}} \\
				\uturnderivcode{\regr_b;} \\
				\uturnderivassertion{\oktext{ok: \exists \vec{a} . s \mapsto \vec{a} \land l = 100 \land pv(2) \land i = 2 \land si = a2}}
					{\oktext{ok: \exists \vec{a} . s \mapsto \vec{a} \land l = 100 \land pv(3) \land i = 2 \land si = a2}} \\
				\uturnderivcode{\regr_b;} \\
				\uturnderivassertion{\oktext{ok: \exists \vec{a} . s \mapsto \vec{a} \land l = 100 \land pv(2) \land i = 3 \land si = a3}}
					{\oktext{ok: \exists \vec{a} . s \mapsto \vec{a} \land l = 100 \land pv(3) \land i = 3 \land si = a3}} \\
				\uturnderivcode{\regr_b;} \\
				\uturnderivassertion{\ertext{er: \exists \vec{a} . s \mapsto \vec{a} \land l = 100 \land pv(3) \land i = 4 \land si = a3}}
					{\ertext{er: \exists \vec{a} . s \mapsto \vec{a} \land l = 100 \land pv(3) \land i = 4 \land si = a3}} \\
				\uturnderivcode{\code{(si == \nulltc)?}} \\
				\uturnderivassertionb{\ertext{er: \exists \vec{a} . s \mapsto \vec{a} \land l = 100 \land pv(3) \land i = 4 \land si = a3}}
					{\ertext{er: \exists \vec{a} . s \mapsto \vec{a} \land l = 100 \land i = 4 \land pv(3)}}
			\end{align*}
			\caption{Sketch of the U-Turn derivation for \code{len}. Following Fig.~\ref{fig:isl-deriv-rlen}, we call $\regr_b$ the body of the while loop and unroll it four times. This is enforced by the proof system since the IL derivation did the same.}
		\end{subfigure}
		\begin{subfigure}[T]{\textwidth}
			\scriptsize
			\begin{align*}
				\uturnderivassertion{\oktext{ok: \true}}
					{\oktext{ok: l = 3}} \\
				\uturnderivcode{\code{s := init(l);}} \\
				\uturnderivassertion{\oktext{ok: \exists \vec{a} . s \mapsto \vec{a} \land l = 3 \land pv(2)}}
					{\oktext{ok: \exists \vec{a} . s \mapsto \vec{a} \land 100 = 100 \land pv(3) \land l = 3}} \\
				\uturnderivcode{\code{l := 100;}} \\
				\uturnderivassertion{\oktext{ok: \exists \vec{a} . s \mapsto \vec{a} \land l = 100 \land pv(2)}}
					{\oktext{ok: \exists \vec{a} . s \mapsto \vec{a} \land l = 100 \land \text{len(s)} = 4 \land pv(3)}} \\
				\uturnderivcode{\code{i := len(s);}} \\
				\uturnderivassertionb{\ertext{er: \exists \vec{a} . s \mapsto \vec{a} \land l = 100 \land pv(3) \land i = 4 \land si = a3}}
					{\ertext{er: \exists \vec{a} . s \mapsto \vec{a} \land l = 100 \land i = 4 \land pv(3)}}
			\end{align*}
			\caption{Sketch of the U-Turn proof for $\mathsf{rlen}$.}
		\end{subfigure}
		\caption{Sketch of the U-Turn derivation for \code{len}. We write it linearized, annotating program points with both the IL and the SIL assertion. The former are the same as Fig.~\ref{fig:isl-deriv-rlen}. The latter are better read bottom-up and form the SIL triple obtained following the IL derivation that lead to the corresponding IL assertions.}
		\label{fig:il+sil-deriv-rlen}
	\end{figure}

	The ISL proof system finds the error: it considers the ``else'' branch of the if statement in \code{init} to find that the code path with $l = 3$ has the string without the null terminator \nullt, that later leads to the error in \code{len} by unrolling the while loop 3 times. However, the ISL triple does not highlight the cause of error, that is the condition $l = 3$ at the beginning of the program.
	
	In theory, Separation SIL can find the source of this error. However, to do so it should guess the right amount of unrolling for the while loop in \code{len}, since there is no indication in the post that $4$ is the number of iterations: this information comes from an earlier program point, that Separation SIL has not explored yet.

	In this example, unity is strength. In fact, the ISL derivation unrolled the while loop exactly $4$ times, because it knew the right number from the condition $l = 3$ in \code{init}. Separation SIL can thus exploit this information: by unrolling the loop 4 times, it finds exactly the error source, that is $l = 3$. This information sharing is formally captured by our combined proof system, whose derivation is shown in Fig.~\ref{fig:il+sil-deriv-rlen}. 
	To help readability, the arrows indicate the order of deductions: first we perform a forward analysis of the code using ISL proof system (the flow of deduction is shown on the left hand side of the figures), which produced the assertions within square brackets, then, once the error is found, we use the proof system of U-Turn to derive SIL triples, i.e., the assertions within angle brackets, by backward analysis (accordingly, the flow of deduction is moved to the right hand side of the figures).
	
	Note the introduction of some constraints in some SIL assertions to ensure that they are subsets of the corresponding ISL assertions. For instance, SIL would not require that $si = a3$ before the assignment \code{si := s[i]} in Fig.~\ref{fig:il+sil-deriv-rlen:1}, or that $l = 3$ before the assignment \code{l := 100}, but since the IL assertions prescribe these additional constraints, they appear in the SIL assertions too. This witnesses another way IL can transfer information to SIL, that we expand in the next example.
\end{example}

\begin{example}\label{ex:uturn-push-back}

\begin{figure}[t]
\hspace{0.1\textwidth}%
\begin{minipage}[t]{0.4\textwidth}
	\begin{lstlisting}[language=C]
// program r
x := [v];
push_back(v);
[x] := 42;
	\end{lstlisting}
\end{minipage}%
\begin{minipage}[t]{0.4\textwidth}
	\begin{lstlisting}[language=C,literate={regplus}{{$\regplus$}}1]
push_back(v) {
  ( y := [v];
    free(y);
    y := alloc();
    [v] := y; )
  regplus ( skip; )
}
	\end{lstlisting}
\end{minipage}
\caption{The program \code{push\_back} discussed in Example~\ref{ex:uturn-push-back}.}
\label{fig:pushback}
\end{figure}

	Consider the \code{push\_back} example in Fig.~\ref{fig:pushback}, already examined in both ISL~\citep{RaadBDDOV20} and Separation SIL~\citep{AscariBGL25} papers.

	Roughly speaking, a ISL analysis can find an error in the assignment \code{[x] := 42} if it picks the left branch in \code{push\_back}, where \code{v} gets reallocated. For Separation SIL to find such an error, it has not only to explore the same branch in \code{push\_back} (the same code path), but also to guess that \code{y} is \emph{aliased} to \code{x}. This (possible) aliasing can be detected automatically \citep[\S 5.5]{AscariBGL25}, but the SIL backwards analysis has no way to know which one is the right choice until earlier in the program (so later in the analysis). Therefore, it must consider both the cases where \code{y} and \code{x} are aliased and when they are not aliased. This is embodied by the disjunction $(x = z \lor x \dealloc{})$ found in the precondition of \code{push\_back} \citep[Fig.~6]{AscariBGL25}. Namely, the computed precondition is made of three disjuncts, corresponding to as many disjunct situations:
	\begin{align*}
		&(\true \andsep v \mapsto z \andsep z \mapsto - \andsep x = z) \lor & \text{reallocation in \code{push\_back}, \code{y} and \code{x} aliased} \\
		&(\true \andsep v \mapsto z \andsep z \mapsto - \andsep x \dealloc) \lor & \text{reallocation in \code{push\_back}, \code{y} and \code{x} distinct}\\
		&(\true \andsep x \dealloc) & \text{no reallocation in \code{push\_back}}
	\end{align*}

	\begin{figure}[t]
	{
		\scriptsize
		\begin{align*}
			\uturnderivassertion{v \mapsto x \andsep x \mapsto -}
				{v \mapsto x \andsep x \mapsto -} \\
			\uturnderivcode{\code{y := [v];}} \\
			\uturnderivassertion{v \mapsto x \andsep x \mapsto - \andsep y = x}
				{v \mapsto x \andsep x \mapsto - \andsep y = x} \\
			\uturnderivcode{\code{free(y);}}\\
			\uturnderivassertion{v \mapsto x \andsep x \dealloc{} \andsep y = x}
				{v \mapsto x \andsep x \dealloc{} \andsep y = x \textcolor{gray}{(\text{instead of } y = x \lor y \dealloc)}} \\
			\uturnderivcode{\code{y := alloc();}} \\
			\uturnderivassertion{v \mapsto x \andsep x \dealloc{} \andsep y \mapsto -}
				{v \mapsto x\textcolor{gray}{(\text{instead of } v \mapsto -)} \andsep x \dealloc{} \andsep y \mapsto -} \\
			\uturnderivcode{\code{[v] := y;}} \\
			\uturnderivassertionb{v \mapsto y \andsep x \dealloc{} \andsep y \mapsto -}
				{v \mapsto y \andsep x \dealloc{} \andsep y \mapsto -}
		\end{align*}
	}
	\caption{Sketch of the U-Turn derivation for \code{push\_back}, linearized. We annotate program points with both the IL and the SIL assertion. The former should be read top down, the latter bottom-up. In the SIL assertions, we write in \textcolor{gray}{gray} what what we would obtain by plain application of the Separation SIL axioms, without the additional constraint to be a stronger assertion than the corresponding IL one.}
	\label{fig:il+sil-deriv-push-back}
	\end{figure}

	Note that only the last line correspond directly to a different program path than the others. However, the U-Turn proof system is able to share enough information between the two analyses to prune also the second disjunct. Intuitively, the ISL analysis contains the information that \code{y} and \code{x} are aliased in the assertions computed during the analysis (in a forward analysis it is easy to know that $v \mapsto x$ already when \code{y := [v]} is executed). The requirement that SIL assertions imply the IL assertions at the same program point forces this information transfer.

	We focus on the left branch of \code{push\_back} only, that we name $\regr_b$. We take the ISL derivation from \citet[Fig.~3]{RaadBDDOV20} and apply U-Turn to it in Fig.~\ref{fig:il+sil-deriv-push-back}. This proves the Separation SIL triple 
	\[
	\siltriple{v \mapsto x \andsep x \mapsto -}{\regr_{b}}{v \mapsto y \andsep x \dealloc{} \andsep y \mapsto -}
	\]
	 as opposed to the triple 
	 \[\siltriple{\true \andsep v \mapsto z \andsep z \mapsto - \andsep (x = z \lor x \dealloc{})}{\regr_{b}}{x \dealloc{} \andsep \true}
	 \] 
	 from \citet[Fig.~6]{AscariBGL25}.
	Particularly, in the triple returned by U-Turn there is only the disjunct where $x = z$, while Separation SIL alone must consider both, cluttering the analysis with useless disjuncts. This is possible due to the information in the IL assertion implicitly flowing into the SIL derivation via the $P' \implies P$ constraint in the soundness of the U-Turn judgment.
\end{example}

\subsection{Progress and Automation}
Since U-Turn must follow the IL derivation closely but imposes additional constraints, it is a non-trivial and practically relevant question whether or not it is always possible to complete a U-Turn proof given any IL derivation $d$.

\begin{lstlisting}[label={fig:u-turn-backwards-alg},language=Haskell,float,extendedchars=true,literate={∃}{{$\exists$}}1 {⊞}{{$\regplus$}}1,caption={Pseudocode of the UTurn algorithm.}]
UTurn :: ILProofTree -> Assertion -> Assertion

-- As stated in the theorem we assume that (false != Q') and (Q' -> Q)

-- Atomic commands
UTurn (ILAssign [P] x:=a [Q]) Q' = (P /\ Q'[a/x])
UTurn (ILAssume [P] b? [Q])   Q' = Q'
UTurn (ILNondet [P] x:=* [Q]) Q' = (P /\ ∃x. Q')
-- Structural rules
UTurn (ILSeq d1 d2 [P] r [Q]) Q' =
  let R' = UTurn d2 Q' in
  let P' = UTurn d1 R' in P'
UTurn (ILErId [P] r [Q]) Q' = Q'
UTurn (ILChoiceL d [P] r1 ⊞ r2 [Q]) Q' = UTurn d Q'
UTurn (ILIter0 [P] r* [Q]) Q' = Q'
UTurn (ILUnroll d [P] r* [Q]) Q' = UTurn d Q'
-- Other rules
UTurn (ILCons d' [P'] r [Q']) Q'' = let P'' = UTurn d' Q'' in P''
UTurn (ILDisj d [P] r [Q]) Q''    =
  let (d1 [P1] r [Q1]), (d2 [P2] r [Q2]) = d in
  let Q1' = Q' /\ Q1, Q2' = Q' /\ Q2 in
  let P1' = if Q1' == False then False else UTurn d1 Q1' in
  let P2' = if Q2' == False then False else UTurn d2 Q2' in
    P1' \/ P2'
\end{lstlisting}

The next theorem not only answers in the affirmative, but also provides a high-level algorithm to do so. We call this algorithm \code{UTurn}, and we present it in Listing~\ref{fig:u-turn-backwards-alg} (we omit cases for rules not in Fig.~\ref{fig:uturn-rules}).
This means that, given any proof tree $d$ for the IL triple $\vdash \iltriple{P}{\regr}{Q}$ and any (non-empty) subset of the errors $Q' \implies Q$, the \code{UTurn} algorithm always yields a $P' \nequiv \false$ such that the judgment $\judgprq$ is provable.
Particularly, given a single error state $\sigma \in Q$, it is always possible to find a non-empty $P'$, i.e., some causes for it, by considering $Q' = \{ \sigma \}$.
Roughly speaking, \code{UTurn} applies the U-Turn rules in a process of backwards inference to find a SIL precondition for any subset of the errors found in IL.

\begin{theorem}\label{th:u-turn-progress}
	Given a derivation $d$ for the IL triple $\vdash \iltriple{P}{\regr}{Q}$ and a $Q'$ such that $\false \nequiv Q' \implies Q$, let $P' = \code{UTurn d Q'}$. Then
	\[
	\judgprq
	\]
\end{theorem}
\begin{proof}[Proof sktech]
	Note that, since the judgment $\judgprq$ is provable and the proof system is sound, then the judgment is valid. Then, point (4) of validity and the hypothesis $Q' \nequiv \false$ imply that $P' \nequiv \false$.

	The proof is by induction on the derivation $d$ of the IL triple.
	Roughly speaking, the proof inspects the last rule applied by the IL triple and applies the homonymous U-Turn rule, always processing the right subtree of \lrule{seq} first. This produces a backward-fashioned derivation, where the post of the current rule is always provided by the previous step, and the pre is computed as prescribed by the applied rule. Note that the use of inductive hypotheses in the proof correspond to recursive calls in the algorithm \code{UTurn}.
\end{proof}

Note that in Examples~\ref{ex:uturn-string-null} and \ref{ex:uturn-push-back} we basically applied the algorithm \code{UTurn} to perform the U-Turn derivations: this is because the algorithm follows naturally from the U-Turn rules, and therefore it gives the most natural (albeit not the only) way to use the proof system.

\subsection{Following SIL Derivations with IL}
In the previous sections, we presented U-Turn for following IL derivations backward with SIL. As anticipated, it is also possible to do the opposite, namely to follow a SIL derivation forward with IL, obtaining a proof system that we informally call Turn-U. For brevity, we do not spell out the rules of Turn-U since they are entirely dual to the one in Fig.~\ref{fig:uturn-rules}, but the corresponding full pseudo-code is available in the Appendix, Listing~\ref{fig:app:u-turn-forwards-alg-full}.
We show below how Turn-U can be useful with an example.

	\begin{figure}[t]
	\begin{align*}
		&\code{foo(b) \{ } \\
		&\quad \code{x := nondet()}; \\
		& \quad\code{if (b $\land$ x $\neq$ 0) \{ p := alloc() \}} \\
		& \quad\code{else \{ p := null \};} \\
		&\quad \code{[p] := x;} \\
		&\quad \code{return p} \\
		&\code{\}}
	\end{align*}
	\caption{The program \code{foo} discussed in Example~\ref{ex:turnu-foo}.}
	\label{fig:foo}
	\end{figure}

\begin{example}\label{ex:turnu-foo}
	Consider the procedure \code{foo(b)} in Fig.~\ref{fig:foo}, 	where we use \code{nondet()} to model some opaque library call for which we do not have any summary available. We want to produce a summary for \code{foo} telling us when it can cause errors, so we start with a SIL analysis from the postcondition \silexact{\ertext{\erflag: \true}}, and we obtain the precondition \silexact{\oktext{\okflag: b}} (some of the intermediate assertions are detailed in the combined derivation below). Unfortunately, this precondition is not informative enough: first, it does not describe precisely which errors can happen in \code{foo} and where; second, it does not say anything about the opaque library call. 
	To fix these two issues, we trace SIL derivation forward using Turn-U. Note that the SIL analysis already found out that the error is caused by the else-branch of the \code{if}, so that IL can analyse only that branch instead of having to check both. Therefore, following the SIL analysis, we obtain the derivation in Fig.~\ref{fig:turnu-derivation}, where we elided the then-branch since it gets ignored:

\begin{figure}[t]
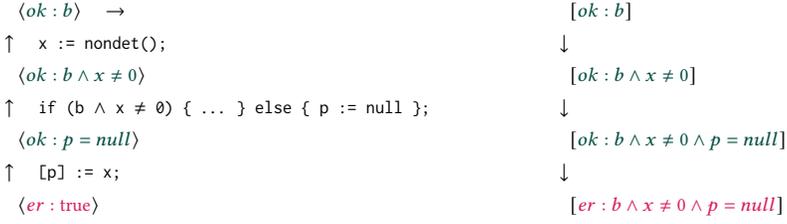

	{ \scriptsize
		\begin{align*}
			\turnuderivassertiont{\oktext{\okflag: b}}
			{\oktext{\okflag: b}} \\
			\turnuderivcode{\code{x := nondet();}} \\
			\turnuderivassertion{\oktext{\okflag: b \land x \neq 0}}
			{\oktext{\okflag: b \land x \neq 0}} \\
			\turnuderivcode{\code{if (b $\land$ x $\neq$ 0) \{ $\dots$ \} else \{ p := null \};}} \\
			\turnuderivassertion{\oktext{\okflag: p = null}}
			{\oktext{\okflag: b \land x \neq 0 \land p = null}} \\
			\turnuderivcode{\code{[p] := x;}} \\
			\turnuderivassertion{\ertext{\erflag: \true}}
			{\ertext{\erflag: b \land x \neq 0 \land p = null}}
		\end{align*}
	}
	\caption{Sketch of the Turn-U proof for the program $\mathsf{foo}$.}
	\label{fig:turnu-derivation}
	\end{figure}
	
	Contrary to previous U-Turn derivation, the flow of deduction is now reversed, as illustrated by the arrows. The analysis started from the postcondition $\ertext{\erflag: \true}$ with the SIL derivation flowing bottom up (on the left hand side of the figure, along the assertions within angle brackets) until the precondition $\oktext{\okflag: b}$ is found and then flow is reversed by Turn-U  (as shown on the right hand side of the figure, along the assertions within square brackets).
	
	Note that the IL postcondition is very informative: not only it includes that $p = null$, the actual cause of the error (the attempt to dereference a null-pointer), but also that $x \neq 0$, therefore giving some information on the result of the (opaque) library call to reach the error. Note that SIL cannot encode it in the precondition because it refers to states \emph{before} the library call and therefore cannot embed any information about the \emph{output} of the call.
\end{example}

\subsection{Relation with UNTer}\label{sec:uturn-il-valid}
We started this section by moving away from the perspective of directly proving triples that are valid for both IL and SIL. This lead us to first prove an IL triple, and then exploit its proof tree to derive a corresponding SIL triple. It turns out we did not stray far from home: if the U-Turn proof does not use SIL consequence rule, \emph{the resulting triple is valid in both IL and SIL}.

\begin{theorem}\label{th:uturn-il-valid}
	If the judgment $\ \judgprq$ is provable without using rule \lrule{consSIL}, then $\vDash \iltriple{P'}{\regr}{Q'}$.
\end{theorem}

We prove this theorem by showing that it is possible to follow again the U-Turn derivation with IL, in the spirit of what we discussed in the previous section. This suggest that it may be possible to obtain the same result even in the presence of SIL consequence rule by executing another forward step to ensure the triple obtained is valid in IL.

We can ensure this by modifying the U-Turn proof system to add a side condition on rule \lrule{consSIL}. The premises of \lrule{consSIL} provide a SIL derivation for the triple $\siltriple{P'}{\regr}{Q'}$, on which we can apply the Turn-U proof system. This will find a triple $\iltriple{P'''}{\regr}{Q'''}$ with $P''' \implies P'$ and $Q''' \implies Q'$. By symmetry with U-Turn, if the Turn-U proof did not use the rule of consequences of IL \rrule{consIL} (a dual of \lrule{consSIL}, where Turn-U can use the IL rule of consequence \ilrule{cons} with some additional constraints), the resulting triple is valid in SIL.
We argue that, in practice, there is no need to use the rule \rrule{consIL} in this latter Turn-U derivation. In fact, the main application of the rule of consequence in under-approximation logics is to drop disjuncts whenever the formulae involved becomes unmanageably large. However, we already know that, for instance, $P''' \implies P' \implies P$, and similarly the assertions at every point of the Turn-U derivation are smaller than the corresponding assertions in the original IL derivation. Therefore, since we already completed the original IL derivation, any formula appearing in it is manageable, so the ones in this Turn-U derivation must be, too. Then, since this Turn-U proof can be carried out without rule \rrule{consIL}, the resulting triple is valid in both IL and SIL and can thus be used to continue the U-Turn derivation, proving a triple that is both IL and SIL in the end.

Note that the algorithm \code{UTurn} never applies \lrule{consSIL}. Therefore, thanks to Theorem~\ref{th:uturn-il-valid}, it always finds triples that are valid both in SIL and IL. The same holds for the algorithm \code{TurnU}. 

This result opens up a direct comparison with UNTer: since both proof systems find triples that are valid both in IL and SIL, why should we prefer one or the other?
There are several points that distinguish the two.

First, UNTer is able to reason about (non)termination of programs, something that U-Turn cannot do. Second, we think the new way of combining triple exemplified by U-Turn is interesting in itself.

On a more technical level, most UNTer rules are inspired by IL in such a way to be applicable to a generic pre (except the rule Assume, only applicable to "specific" pre and post, cf. Section~\ref{sec:unter-bg}). This means that they are not immediately applicable (algorithmically, for backward analysis) to generic post-conditions, unlike U-Turn rules. As a downside, UNTer works in a single-pass algorithm, while U-Turn requires a first forward, IL-based step followed by a second backwards, SIL-based step, leading to a two-passes algorithm.

When considering completeness of the two approaches there is no clear winner either.
On the one hand, UNTer has no completeness result for triples that valid in both IL and SIL at the same time.
On the other hand, U-Turn is incomplete as well, in the traditional sense of the word (that is, every valid judgment is provable).
However, nailing down the right notion of completeness for U-Turn is not a straightforward task. The above meaning of completeness is a very strong property: namely, given any derivation $d$ for an IL triple $\vdash \iltriple{P}{\regr}{Q}$ and any valid SIL triple $\vDash \siltriple{P'}{\regr}{Q'}$ such that $P' \implies P$ and $Q' \implies Q$, the resulting U-Turn judgment is provable.
A different but still relevant notion would be to require, for any provable IL triple $\vdash \iltriple{P}{\regr}{Q}$ and valid SIL triple $\vDash \siltriple{P'}{\regr}{Q'}$, the \emph{existence} of a proof tree $d$ for $\vdash \iltriple{P}{\regr}{Q}$ making the U-Turn judgment provable. Currently, we are still unsure whether this is the case. We leave this question as future work.

\section{Conclusions}\label{sec:conclusions}

Postconditions of triples in Incorrectness Logic---a forward under-approximation analysis---only expose errors that are reachable from the preconditions, but not all initial states described  by the pre necessarily lead to errors. 
 Conversely, the preconditions of triples in Sufficient Incorrectness Logic—a backward under-approximation analysis—include only initial states that can cause some of the errors in the postcondition, but not necessarily all of them.
When the same triple is valid in both logics we have the best of both worlds: we are guaranteed that any initial state in the pre is the source of some error in the post and that all errors in the post are reachable from states in the pre.
This form of summaries provides a highly valuable feedback for developers, because there are no false alarms and the sources of errors can facilitate  testing and debugging activities.

In this paper we have explored the combination of forward and backward under-approximation approaches, to improve the precision of the analysis and to be able to match reachable errors with their sources. 
We are not the first ones to consider such a combined analysis: for instance, UNTer logic already considered under-approximation triples that are valid in both directions, although, in that case, the emphasis is on exploiting backward under-approximation triples for non-termination analysis~\cite{RaadVO24}. 
We advance the state-of-the-art on this combination in two ways.
First, the axioms for atomic commands in UNTer are handcrafted starting from the availability of a tool for forward analysis, namely PulseX. Therefore they required human ingenuity and are difficult to extend with new primitives.
Moreover, they cannot be applied to either generic preconditions or postconditions. 
In this respect, we have provided a methodology to derive axiom schemes that are sound and complete by construction, thus solving all the above issues.
Second, we propose a new way to combine IL and SIL analyses: rather than trying to derive directly a triple valid in both logics, we suggest a clever  composition of the two techniques, where after deriving a valid triple in one logic we refine the proof to get a triple valid in both logics in an automatic way.
Following these ideas, we defined a novel proof system, called  U-Turn, whose judgments present a novel shape to compose derivations in different logics.
Our main results show that U-Turn is sound, that it is able to derive triples valid in both SIL and IL under suitable assumption, and show how this inference can be automated.
Interestingly, U-Turn can be used to refine a preliminary IL analysis using SIL or vice versa.
Moreover, we have shown that whenever some additional form of approximation is necessary in one direction, e.g., to improve the performance of the analysis by dropping further disjuncts, the two ways of invoking U-Turn can mutually cooperate.

\paragraph*{Future Works.}
There are many interesting directions we plan to explore further.

First, we want to investigate the question of U-Turn completeness relative to different notions, both on a technical side and by understanding the meaningfulness of such definition with respect to applications.

Second, the shape of U-Turn judgments open the possibility to investigate the combination of different proof systems by reusing a derivation in one logic to drive the inference in the other logic. We think it would be interesting to see how far this principle can be extended to combination of over- and under-approximations.

Third, the use of under-approximation approaches to incorrectness reasoning has already been paired with abstract interpretation techniques for correctness analysis, and we would like to extend U-Turn in this respect. 
In \citet{BruniGGR21} the authors introduce Local Completeness Logic to combine the derivation of IL triples $\iltriple{P}{\regr}{Q}$ with over-approximation in an abstract domain $A$ in order to guarantee that the states that are reachable from $P$ executing $\regr$ is, at the same time, under-approximated by $Q$ (if an error is in $Q$ it is reachable and must be reported to developers) and over-approximated by the best abstraction of $Q$ in the domain $A$, denoted $A(Q)$  (if $A(Q)$ is bug free, then the program is correct). The local completeness condition further guarantees that  $A(Q)$ is bug free if and only if $Q$ is bug free, so that the same under-approximation $Q$ can be used for both correctness and incorrectness reasoning.
We are confident that the local completeness technique can be smoothly extended to backward under-approximation and possibly integrated with U-Turn so that at least one source of errors will be exposed by the analysis whenever some error is possible.

Fourth, the growing interest around hyperlogics for studying relational and hyperproperties~\cite{DBLP:conf/popl/Benton04,DBLP:conf/csfw/ClarksonS08,DBLP:conf/pldi/SousaD16,DBLP:journals/pacmpl/DardinierM24,DBLP:journals/pacmpl/CousotW25} can provide some challenging analysis scenarios, where U-Turn approach can play a fundamental role in tackling the complexity of the state-space by means of under-approximation.

Finally, we mention the possibility to extend and integrate bi-abduction techniques for backward reasoning within U-Turn derivation, to ease their scalability and reuse in the analysis of industrial size code.

\bibliographystyle{ACM-Reference-Format}
\bibliography{bibfile}

\appendix

\newpage
\section{Proofs}\label{sec:proofs}

\begin{proof}[Proof of Lemma~\ref{lmm:assignment-fw-bw}]
	This Lemma is a corollary of Lemma~4.8 in \citet{AscariBGL25} by observing that assignments are total, that is (in the notation of the aforementioned Lemma~4.8) $D_{\code{x:=a}} = \emptyset$ 
\end{proof}

\begin{proof}[Proof of Lemma~\ref{lmm:p-q-valid-empty}]
	Validity of the triple $\vDash \siltriple{P}{\regr}{Q}$ corresponds to $P \subseteq \bwsem{\regr} Q = \bwsem{\regr} \emptyset = \emptyset$.
	
	Validity of the triple $\vDash \iltriple{P}{\regr}{Q}$ corresponds to $Q \subseteq \fwsem{\regr} P = \fwsem{\regr} \emptyset = \emptyset$.
\end{proof}

\begin{proof}[Proof of Proposition~\ref{prop:axiom-schema-valid}]
	We prove only the validity of the triple in IL, the proof for SIL is analogous.
	
	Take a state $\sigma' \in Q \land \fwsem{\regc} P$.	By definition of $\fwsem{\regc}$, since $\sigma' \in \fwsem{\regc} P$ there exists a $\sigma \in P$ such that $\sigma' \in \fwsem{\regc} \sigma$. We then show that $\sigma \in \bwsem{\regc} Q$.
	We know that $\sigma' \in Q$ and $\sigma' \in \fwsem{\regc} \sigma$, that is equivalent to $\sigma \in \bwsem{\regc} \sigma' \subseteq \bwsem{\regc} Q$.
	This proves validity in IL: given any state $\sigma'$ in the post, there exists a state $\sigma$ in the pre that can reach it.
\end{proof}

\begin{proof}[Proof of Proposition~\ref{prop:axiom-schema-complete}]
	By hypothesis, $\vDash \iltriple{P}{\regc}{Q}$, which means that $Q \subseteq \fwsem{\regc} P$. Therefore, $Q \land \fwsem{\regc} P = Q$.
	The proof for $P \land \bwsem{\regc} Q = P$ is analogous using the hypothesis that $\vDash \siltriple{P}{\regc}{Q}$ instead.
\end{proof}

\subsection{U-Turn Proof System and Soundness}

\begin{figure}[h]
	\centering
	\begin{framed}
		\scalebox{0.85}{
			\(
			\begin{array}{c}
				\infer[\lrule{assign}]
				{ \ilsiljudgment{\ilderiv{\phantom{d}}{\oktext{ok: P}}{\code{x := a}}{\oktext{ok: \spa{x}{a}{P}}}}{\oktext{ok: P \land Q'[a / x]}}{\code{x := a}}{\oktext{ok: Q'}}}
				{ Q' \implies \spa{x}{a}{P} }
				\\[4.5ex]
				\infer[\lrule{nondet}]
				{ \ilsiljudgment{\ilderiv{\phantom{d}}{\oktext{ok: P}}{\code{x := nondet()}}{\oktext{ok: \exists x . P}}}{\oktext{ok: P \land \exists x . Q'}}{\code{x := nondet()}}{\oktext{ok: Q'}} }
				{ Q' \implies \exists x . P }
				\\[4.5ex]
				\infer[\lrule{assume}]
				{ \ilsiljudgment{\ilderiv{\phantom{d}}{\oktext{ok: P}}{\code{b?}}{\oktext{ok: P \land b}}}{\oktext{ok: Q'}}{\code{b?}}{\oktext{ok: Q'}} }
				{ Q' \implies P \land b }
				\\[4.5ex]
				\infer[\lrule{skip}]
				{ \ilsiljudgment{\ilderiv{\phantom{d}}{P}{\code{skip}}{P}}{Q'}{\code{skip}}{Q'} }
				{ Q' \implies P }
				\qquad
				\infer[\lrule{er\mbox{-}id}]
				{ \ilsiljudgment{ \ilderiv{\phantom{d}}{\ertext{er: P}}{\regr}{\ertext{er: P}} }{\ertext{er: Q'}}{\regr}{\ertext{er: Q'}} }
				{ Q' \implies P }
				\\[4.5ex]
				\infer[\lrule{empty}]
				{ \ilsiljudgment{\ilderiv{d}{P}{\regr}{Q}}{\false}{\regr}{\false} }
				{ }
				\\[4.5ex]
				\infer[\lrule{disj}]
				{ \ilsiljudgment{\ilderiv{\ilderiv{d_1}{P_1}{\regr}{Q_1} & \ilderiv{d_2}{P_2}{\regr}{Q_2}}{P_1 \lor P_2}{\regr}{Q_1 \lor Q_2}}{P_1' \lor P_2'}{\regr}{Q_1' \lor Q_2'} }
				{ \ilsiljudgment{\ilderiv{d_1}{P_1}{\regr}{Q_1}}{P_1'}{\regr}{Q_1'} & \ilsiljudgment{\ilderiv{d_2}{P_2}{\regr}{Q_2}}{P_2'}{\regr}{Q_2'} }
				\\[6ex]
				\infer[\lrule{seq}]
				{ \ilsiljudgment{\ilderiv{\ilderiv{d_1}{P}{\regr_1}{R} & \ilderiv{d_2}{R}{\regr_2}{Q}}{P}{\regr_1; \regr_2}{Q}}{P'}{\regr_1; \regr_2}{Q'} }
				{ \ilsiljudgment{\ilderiv{d_1}{P}{\regr_1}{R}}{P'}{\regr_1}{R'} & \ilsiljudgment{\ilderiv{d_2}{R}{\regr_2}{Q}}{R'}{\regr_2}{Q'} }
			\end{array}
			\)
		}
	\end{framed}
	\Description{See caption.}
	\caption[]{The complete U-Turn proof system (part 1).}
\end{figure}

\begin{figure}[h]
	\ContinuedFloat
	\centering
	\begin{framed}
		\scalebox{0.85}{
			\(
			\begin{array}{c}
				\infer[\lrule{choiceL}]
				{ \ilsiljudgment{\ilderiv{\ilderiv{d}{P}{\regr_1}{Q}}{P}{\regr_1 \regplus \regr_2}{Q}}{P'}{\regr_1 \regplus \regr_2}{Q'} }
				{ \ilsiljudgment{\ilderiv{d}{P}{\regr_1}{Q}}{P'}{\regr_1}{Q'} }
				\quad %
				\infer[\lrule{choiceR}]
				{ \ilsiljudgment{\ilderiv{\ilderiv{d}{P}{\regr_2}{Q}}{P}{\regr_1 \regplus \regr_2}{Q}}{P'}{\regr_1 \regplus \regr_2}{Q'} }
				{ \ilsiljudgment{\ilderiv{d}{P}{\regr_2}{Q}}{P'}{\regr_2}{Q'} }
				\\[5ex]
				\infer[\lrule{iter0}]
				{ \ilsiljudgment{\ilderiv{\phantom{d}}{P}{\regr^{\kstar}}{P}}{Q'}{\regr^{\kstar}}{Q'} }
				{ Q' \implies P }
				\qquad\qquad %
				\infer[\lrule{unroll}]
				{ \ilsiljudgment{\ilderiv{\ilderiv{d}{P}{\regr^{\kstar}; \regr}{Q}}{P}{\regr^{\kstar}}{Q}}{P'}{\regr^{\kstar}}{Q'} }
				{ \ilsiljudgment{\ilderiv{d}{P}{\regr^{\kstar}; \regr}{Q}}{P'}{\regr^{\kstar}; \regr}{Q'} }
				\\[5ex]
				\infer[\lrule{consIL}]
				{ \ilsiljudgment{\ilderiv{P' \implies P & \ilderiv{d}{P'}{\regr}{Q'} & Q \implies Q'}{P}{\regr}{Q}}{P''}{\regr}{Q''} }
				{ \ilsiljudgment{\ilderiv{d}{P'}{\regr}{Q'}}{P''}{\regr}{Q''} & Q'' \implies Q }
				\\[4.5ex]
				\infer[\lrule{consSIL}]
				{ \ilsiljudgment{\ilderiv{d}{P}{\regr}{Q}}{P'}{\regr}{Q'} }
				{ \false \nequiv P' \implies P'' & \ilsiljudgment{\ilderiv{d}{P}{\regr}{Q}}{P''}{\regr}{Q''} & Q'' \implies Q' \implies Q }
			\end{array}
			\)
		}
	\end{framed}
	\Description{See caption.}
	\caption{The complete U-Turn proof system (part 2).}\label{fig:app:uturn-rules-full}
\end{figure}

\begin{lemma}\label{lmm:assign-valid-for-il}
	If
	\[
	\infer
	{ \ilsiljudgment{\ilderiv{\phantom{d}}{\oktext{ok: P}}{\code{x := a}}{\oktext{ok: \spa{x}{a}{P}}}}{\oktext{ok: P \land Q'[a / x]}}{\code{x := a}}{\oktext{ok: Q'}}}
	{ \false \nequiv Q' \implies \spa{x}{a}{P} }
	\]
	then $\vdash \iltriple{\oktext{ok: P \land Q'[a / x]}}{\code{x := a}}{\oktext{ok: Q'}}$.
\end{lemma}
\begin{proof}
	By using \ilrule{assign} we prove the IL triple
	\[
	\vdash \iltriple{\oktext{ok: P \land Q'[a / x]}}{\code{x := a}}{\oktext{ok: \exists y' . P[y' / x] \land Q'[a / x][y' / x] \land x = a[y' / x]}}
	\]
	
	Since $y'$ is fresh with respect to $P \land Q'[a / x]$ and $x$, it doesn't appear in $Q'$ either. Moreover, $Q'[a / x][y' / x] = Q'[a[y' / x] / x]$ because all other occurrences of $x$ in $Q'$ other than the ones in $a$ are replaced by the first substitution $[a / x]$.
	With these, we derive the following chain of logical equivalences:
	\begin{align*}
		&\quad\; \exists y' . P[y' / x] \land Q'[a / x][y' / x] \land x = a[y' / x] &[\text{observed above}] \\
		&\equiv \exists y' . P[y' / x] \land Q'[a[y' / x] / x] \land x = a[y' / x] &[x = a[y' / x]] \\
		&\equiv \exists y' . P[y' / x] \land Q' \land x = a[y' / x] &[y' \text{ not free in } Q'] \\
		&\equiv (\exists y' . P[y' / x] \land x = a[y' / x]) \land Q' &[\text{def of } \spa{x}{a}{P}] \\
		&\equiv \spa{x}{a}{P} \land Q' &[Q' \implies \spa{x}{a}{P}] \\
		&\equiv Q'
	\end{align*}
	
	Therefore we have $\vdash \iltriple{\oktext{ok: P \land Q'[a / x]}}{\code{x := a}}{\oktext{ok: Q'}}$
\end{proof}

\begin{lemma}\label{lmm:nondet-valid-for-il}
	If
	\[
	\infer
	{ \ilsiljudgment{\ilderiv{\phantom{d}}{\oktext{ok: P}}{\code{x := nondet()}}{\oktext{ok: \exists x . P}}}{\oktext{ok: P \land \exists x . Q'}}{\code{x := nondet()}}{\oktext{ok: Q'}}}
	{ Q' \implies \exists x . P }
	\]
	then $\vdash \iltriple{\oktext{ok: P \land \exists x . Q'}}{\code{x := nondet()}}{\oktext{ok: Q'}}$.
\end{lemma}
\begin{proof}
	By using \ilrule{nondet} we prove the IL triple
	\[
	\vdash \iltriple{\oktext{ok: P \land \exists x . Q'}}{\code{x := nondet()}}{\oktext{ok: \exists x . (P \land \exists x . Q')}}
	\]
	
	With then derive the following chain of logical implications:
	\begin{align*}
		&Q'  &[Q' \implies \exists x . Q'] \\
		\equiv\, &Q' \land \exists x . Q' &[\text{hp of the rule}] \\
		\implies &\exists x . P \land \exists x . Q' &[x \notin \fv(\exists x . Q')] \\
		\equiv\, &\exists x . (P \land \exists x . Q') &
	\end{align*}
	
	Therefore, using this implication and rule \ilrule{cons} we can prove the triple:
	
	\[
	\infer[\ilrule{cons}]
	{\vdash \iltriple{\oktext{ok: P \land Q'[a / x]}}{\code{x := nondet()}}{\oktext{ok: Q'}}}
	{
		\infer[\ilrule{nondet}]
		{\vdash \iltriple{\oktext{ok: P \land \exists x . Q'}}{\code{x := nondet()}}{\oktext{ok: \exists x . (P \land \exists x . Q')}}}
		{}
	}
	\]
\end{proof}

\begin{proof}[Proof of Theorem~\ref{th:uturn-soundness}]
	The proof is by induction on the derivation tree. We also observe that validity condition (4) can be equivalently rewritten as $(P' \equiv \false) \iff (Q' \equiv \false)$. Moreover, by validity of the SIL triple $\vDash \siltriple{P'}{\regr}{Q'}$ and Lemma~\ref{lmm:p-q-valid-empty}.1, one of the implications is already proved, so we only need to show $(P' \equiv \false) \implies (Q' \equiv \false)$.
	
	\proofcase{\lrule{assign}}
	Validity of $\vDash \siltriple{\oktext{ok: P \land Q'[a / x]}}{\code{x := a}}{\oktext{ok: Q'}}$ follows from the SIL derivation below and soundness of the SIL proof system:
	\[
	\infer[\silrule{cons}]
	{\vdash \siltriple{\oktext{ok: P \land Q'[a / x]}}{\code{x := a}}{\oktext{ok: Q'}}}
	{\oktext{ok: P \land Q'[a / x]} \implies \oktext{ok: Q'[a / x]} & \infer[\silrule{assign}]{\vdash \siltriple{\oktext{ok: Q'[a / x]}}{\code{x := a}}{\oktext{ok: Q'}}}{}}
	\]
	
	By hypothesis of the rule, $\oktext{ok: Q'} \implies \oktext{ok: \spa{x}{a}{P}}$. Trivially, $\oktext{ok: P \land Q'[a / x]} \implies \oktext{ok: P}$.
	Lastly, suppose $\oktext{ok: P \land Q'[a / x]} \equiv \false$. By Lemma~\ref{lmm:assign-valid-for-il}, $\vdash \iltriple{\oktext{ok: P \land Q'[a / x]}}{\code{x := a}}{\oktext{ok: Q'}}$. By soundness of the IL proof system, this implies $\vDash \iltriple{\oktext{ok: P \land Q'[a / x]}}{\code{x := a}}{\oktext{ok: Q'}}$. Then, by Lemma~\ref{lmm:p-q-valid-empty}.2, we get $\oktext{ok: P \land Q'[a / x]} \equiv \false$, as desired.
	
	\proofcase{\lrule{nondet}}
	The proof follows the same line as the assign case, using Lemma~\ref{lmm:nondet-valid-for-il}.

	\proofcase{\lrule{assume}}
	The proof is analogous to the assign case.
	
	\proofcase{\lrule{skip}}
	The proof is analogous to the assign case.
	
	\proofcase{\lrule{er\mbox{-}id}}
	Validity of $\vDash \siltriple{\ertext{er: Q'}}{\regr}{\ertext{er: Q'}}$ follows from SIL rule $\silrule{er\mbox{-}id}$, and $\ertext{er: Q'} \implies \ertext{er: P}$ by hypothesis of the rule.
	Lastly, if $(P' \equiv \false)$ then $(Q' \equiv \false)$ because $Q' \implies P \equiv \false$.
	
	\proofcase{\lrule{empty}}
	Validity of $\vDash \siltriple{\false}{\regr}{\false}$ follows from SIL rule $\silrule{empty}$ instantiated with postcondition $Q = \false$.
	We trivially have that $\false \implies P$ and $\false \implies Q$. Lastly, $(\false \equiv \false) \implies (\false \equiv \false)$ concludes the proof.
	
	\proofcase{\lrule{disj}}
	Validity of $\vDash \siltriple{P_1' \lor P_2}{\regr}{Q_1' \lor Q_2'}$ follows from SIL rule $\silrule{disj}$ and inductive hypothesis.
	The implication $P_1' \lor P_2' \implies P_1 \lor P_2$ follows from the two implications $P_i' \implies P_i$, obtained from inductive hypotheses, point (2), on the derivable judgments $\ilsiljudgment{\ilderiv{d_i}{P_i}{\regr}{Q_i}}{P_i'}{\regr}{Q_i'}$. Analogously for $Q_1' \lor Q_2' \implies Q_1 \lor Q_2$ using point (3).
	Lastly, suppose $P_1' \lor P_2' \equiv \false$. Therefore, both $P_i' \equiv \false$. By validity of the judgments $\ilsiljudgment{\ilderiv{d_i}{P_i}{\regr}{Q_i}}{P_i'}{\regr}{Q_i'}$ we then obtain $Q_i' \equiv \false$ from point (4), so that $Q_1' \lor Q_2' \equiv \false$.

	\proofcase{\lrule{seq}}
	Validity of $\vDash \siltriple{P}{\regr_1; \regr_2}{Q}$ follows from SIL rule $\silrule{seq}$ and inductive hypothesis.
	The implications $P' \implies P$ follows from inductive hypotheses on the derivable judgment $\ilsiljudgment{\ilderiv{d_1}{P}{\regr_1}{R}}{P'}{\regr_1}{R'}$, point (2). Analogously for $Q' \implies Q$ with the judgment $\ilsiljudgment{\ilderiv{d_2}{R}{\regr_2}{Q}}{R'}{\regr_2}{Q'}$, point (3).
	Lastly, from point (4) of validity of those two judgments we get $(P' \equiv \false) \iff (R' \equiv \false)$ and $(R' \equiv \false) \iff (Q' \equiv \false)$, so that point (4) of the validity of the derived judgment follows.
	
	\proofcase{\lrule{choiceL}}
	By inductive hypothesis, the judgment $\ilsiljudgment{\ilderiv{d}{P}{\regr_1}{Q}}{P'}{\regr_1}{Q'}$ is valid. Hence, $\vDash \siltriple{P'}{\regr_1}{Q'}$, $P' \implies P$, $Q' \implies Q$ and $(P' \equiv \false) \iff (Q' \equiv \false)$.
	Validity of $\vDash \siltriple{P'}{\regr_1 \regplus \regr_2}{Q'}$ follows from validity of $\vDash \siltriple{P'}{\regr_1}{Q'}$ and SIL rule $\silrule{choiceL}$. The other conditions are exactly given by the inductive hypothesis.
	
	\proofcase{\lrule{chioceR}, \lrule{iter0}, \lrule{unroll}}
	The proof is analogous to the choiceL case.
	
	\proofcase{\lrule{consIL}}
	By inductive hypothesis, the judgment $\ilsiljudgment{\ilderiv{d}{P'}{\regr}{Q'}}{P''}{\regr}{Q''}$ is valid. Hence, $\vDash \siltriple{P''}{\regr}{Q''}$, $P'' \implies P'$, $Q'' \implies Q'$ and $(P'' \equiv \false) \iff (Q'' \equiv \false)$.
	To conclude the proof of the inductive case, we only need to show that $P'' \implies P$ and $Q'' \implies Q$. The latter is an hypothesis of the rule. The former follows from validity of the judgment and the hypothesis $P' \implies P$ of the IL triple.
	
	\proofcase{\lrule{consSIL}}
	By inductive hypothesis, the judgment $\ilsiljudgment{\ilderiv{d}{P}{\regr}{Q}}{P''}{\regr}{Q''}$ is valid. Hence, $\vDash \siltriple{P''}{\regr}{Q''}$, $P'' \implies P$ and $Q'' \implies Q$.
	Validity of $\siltriple{P'}{\regr}{Q'}$ follows from $\vDash \siltriple{P''}{\regr}{Q''}$, the two hypotheses of the rule $P' \implies P''$ and $Q'' \implies Q'$ and SIL rule $\silrule{cons}$.
	$P' \implies P''$ by hypothesis of the rule, and by validity of the judgment $P'' \implies P$, so $P' \implies P$.
	$Q' \implies Q$ by hypothesis of the rule.
	Lastly, we have to show that $(P' \equiv \false) \implies (Q' \equiv \false)$. However, by hypothesis of the rule  $P' \nequiv \false$, so this implication is vacuously satisfied.
\end{proof}

\begin{lstlisting}[label={fig:app:u-turn-backwards-alg-full},language=Haskell,float,extendedchars=true,literate={∃}{{$\exists$}}1 {⊞}{{$\regplus$}}1,caption={Pseudocode of the UTurn algorithm.}]
UTurn :: ILProofTree -> Assertion -> Assertion

-- As stated in the theorem we assume that (false != Q') and (Q' -> Q)

-- Atomic commands
UTurn (ILAssign [P] x:=a [Q]) Q' = (P /\ Q'[a/x])
UTurn (ILAssume [P] b? [Q])   Q' = Q'
UTurn (ILNondet [P] x:=* [Q]) Q' = (P /\ ∃x. Q')
UTurn (ILSkip [P] skip [P])   Q' = Q'
-- Structural rules
UTurn (ILSeq d1 d2 [P] r [Q]) Q' =
  let R' = UTurn d2 Q' in
  let P' = UTurn d1 R' in P'
UTurn (ILErId [P] r [Q]) Q' = Q'
UTurn (ILChoiceL d [P] r1 ⊞ r2 [Q]) Q' = UTurn d Q'
UTurn (ILChoiceR d [P] r1 ⊞ r2 [Q]) Q' = UTurn d Q'
UTurn (ILIter0 [P] r* [Q]) Q' = Q'
UTurn (ILUnroll d [P] r* [Q]) Q' = UTurn d Q'
-- Other rules
UTurn (ILCons d' [P'] r [Q']) Q'' = let P'' = UTurn d' Q'' in P''
UTurn (ILDisj d [P] r [Q]) Q''    =
  let (d1 [P1] r [Q1]), (d2 [P2] r [Q2]) = d in
  let Q1' = Q' /\ Q1, Q2' = Q' /\ Q2 in
  let P1' = if Q1' == False then False else UTurn d1 Q1' in
  let P2' = if Q2' == False then False else UTurn d2 Q2' in
    P1' \/ P2'
\end{lstlisting}

\begin{lstlisting}[label={fig:app:u-turn-forwards-alg-full},language=Haskell,float,extendedchars=true,literate={∃}{{$\exists$}}1 {⊞}{{$\regplus$}}1,caption={Pseudocode of the TurnU algorithm.}]
TurnU :: SILProofTree -> Assertion -> Assertion

-- We assume that (false != P') and (P' -> P)

-- Atomic commands
TurnU (SILAssign <P> x:=a <Q>) P' = (Q /\ ∃x'.P[x'/x] /\ x=a[x'/x])
TurnU (SILAssume <P> b? <Q>)   P' = P'
TurnU (SILNondet <P> x:=* <Q>) P' = (Q /\ ∃x. P')
TurnU (SILSkip <P> skip <P>)   P' = P'
-- Structural rules
TurnU (SILSeq d1 d2 <P> r <Q>) P' =
  let R' = TurnU d1 P' in
  let Q' = TurnU d2 R' in Q'
TurnU (SILChoiceL d <P> r1 ⊞ r2 <Q>) P' = TurnU d P'
TurnU (SILChoiceR d <P> r1 ⊞ r2 <Q>) P' = TurnU d P'
TurnU (SILIter0 <P> r* <Q>) P' = P'
TurnU (SILUnroll d <P> r* <Q>) P' = TurnU d P'
-- Other rules
TurnU (SILCons d' <P'> r <Q'>) P'' = let Q'' = TurnU d' P'' in Q''
TurnU (SILDisj d <P> r <Q>) P''    =
  let (d1 <P1> r <Q1>), (d2 <P2> r <Q2>) = d in
  let P1' = P' /\ P1, P2' = P' /\ P2 in
  let Q1' = if P1' == False then False else TurnU d1 P1' in
  let Q2' = if P2' == False then False else TurnU d2 P2' in
    Q1' \/ Q2'
\end{lstlisting}

\begin{proof}[Proof of Theorem~\ref{th:u-turn-progress}]
	The proof is by induction on the derivation $d$ of the IL triple.

	\proofcase{\ilrule{assign}}
	By hypothesis of the theorem, $Q' \implies Q = \oktext{ok: \spa{x}{a}{P}}$. Therefore, if we take $P' = \oktext{ok: P \land Q'[a / x]} = \code{UTurn d Q'}$ we have that the triple is provable using \lrule{assign}.

	\proofcase{\ilrule{assume}, \ilrule{skip}}
	In both cases, we can take $P' = \oktext{ok: Q'} = \code{UTurn d Q'}$ and prove the U-Turn triple with the homonymous U-Turn rule. The side conditions to apply the U-Turn rules are ensured by the hypothesis of the theorem $Q' \implies Q$.
	
	\proofcase{\ilrule{nondet}}
	By hypothesis, $Q' \implies Q = \oktext{ok: \exists x . P}$. Therefore, if we take $P' = \oktext{ok: P \land \exists x . Q'} = \code{UTurn d Q'}$ we have that the triple is provable using \lrule{nondet}.

	\proofcase{\ilrule{disj}}
	Define $Q_1' = Q' \land Q_1$ and $Q_2' = Q' \land Q_2$. We have that
	\[
	Q_1' \lor Q_2' \equiv (Q' \land Q_1) \lor (Q' \land Q_2) \equiv Q' \land (Q_1 \lor Q_2) \equiv Q'
	\]
	where the last equivalence follows from the hypothesis that $Q' \implies Q_1 \lor Q_2$. Since by hypothesis $\false \nequiv Q'$, at least one of $Q_1'$ and $Q_2'$ is not equivalent to $\false$. Without loss of generality, we can assume $Q_1' \nequiv \false$. One of the hypothesis of the IL triple is that $\iltriple{P_1}{\regr}{Q_1}$ is provable with some derivation $d_1$, and we know $\false \nequiv Q_1' \implies Q_1$. Therefore, by inductive hypothesis, we have a $P_1'$ such that $\ilsiljudgment{\ilderiv{d_1}{P_1}{\regr}{Q_1}}{P_1'}{\regr}{Q_1'}$.
	We now distinguish two cases. If $Q_2' \nequiv \false$, analogously to what we did for $Q_1'$ we get a $P_2'$ such that $\ilsiljudgment{\ilderiv{d_2}{P_2}{\regr}{Q_2}}{P_2'}{\regr}{Q_2'}$.
	If instead $Q_2' \equiv \false$, we take $P_2' = \false$ and use \lrule{empty} to prove $\ilsiljudgment{\ilderiv{d_2}{P_2}{\regr}{Q_2}}{\false}{\regr}{\false}$.
	
	Then, we define $P' = P_1' \lor P_2'$ and combine the two triples using \lrule{disj}:
	\[
	\infer[\lrule{disj}]
	{ \ilsiljudgment{ \infer[\ilrule{disj}]{ \iltriple{P_1 \lor P_2}{\regr}{Q_1 \lor Q_2} }{\ilderiv{d_1}{P_1}{\regr}{Q_1} & \ilderiv{d_2}{P_2}{\regr}{Q_2}} }{P'}{\regr}{Q'} }
	{ \ilsiljudgment{ \ilderiv{d_1}{P_1}{\regr}{Q_1} }{P_1'}{\regr}{Q_1'} & \ilsiljudgment{ \ilderiv{d_2}{P_2}{\regr}{Q_2} }{P_2'}{\regr}{Q_2'} }
	\]
	
	Note that $Q_1'$, $Q_2'$, $P_1'$ and $P_2'$ are precisely as defined in the \code{ILDisj} case of $\code{UTurn}$, which in the end returns exactly $P_1' \lor P_2' = P'$.

	\proofcase{\ilrule{cons}}
	Because of name clashes, following rule \lrule{consIL} in Fig.~\ref{fig:uturn-rules}, we rename to $Q''$ the assertion $Q'$ from the statement of the Theorem.
	One of the hypothesis of the IL triple is that $\iltriple{P'}{\regr}{Q'}$ is provable with some derivation $d'$. By hypothesis, we know $\false \nequiv Q'' \implies Q$, and by hypothesis of the IL triple $Q \implies Q'$, so that $Q'' \implies Q'$. Therefore, we can apply the inductive hypothesis on $\iltriple{P'}{\regr}{Q'}$ and its derivation $d'$ with postcondition $Q''$ to obtain a $P''$ such that $\ilsiljudgment{ \ilderiv{d'}{P'}{\regr}{Q'} }{P''}{\regr}{Q''}$. Particularly, $\code{UTurn d' Q''}$ computes one such $P''$. Again, by soundness (Theorem~\ref{th:uturn-soundness}) also $\false \nequiv P''$, and we can conclude the inductive case with \lrule{consIL}
	\[
	\infer[\lrule{consIL}]
	{ \ilsiljudgment{\ilderiv{P' \implies P & \ilderiv{d'}{P'}{\regr}{Q'} & Q \implies Q'}{P}{\regr}{Q}}{P''}{\regr}{Q''} }
	{ \ilsiljudgment{\ilderiv{d'}{P'}{\regr}{Q'}}{P''}{\regr}{Q''} & Q'' \implies Q }
	\]

	\proofcase{\ilrule{seq}}
	One of the hypothesis of the IL triple is that $\iltriple{R}{\regr_2}{Q}$ is provable with some derivation $d_2$. By inductive hypothesis, we then get a $R' = \code{UTurn d2 Q'}$ such that $\ilsiljudgment{\ilderiv{d_2}{R}{\regr_2}{Q}}{R'}{\regr_2}{Q'}$. Since by hypothesis $Q' \nequiv \false$, by soundness (Theorem~\ref{th:uturn-soundness}) also $\false \nequiv R' \implies R$. Therefore, we can apply the inductive hypothesis on $\iltriple{P}{\regr_1}{R}$ and its derivation $d_1$, that are the other hypothesis of the IL triple. Thus, we get $P' = \code{UTurn d1 R'}$ such that $\ilsiljudgment{\ilderiv{d_1}{P}{\regr_1}{R}}{P'}{\regr_1}{R'}$.
	We can then combine the two triples using \lrule{seq}:
	\[
	\infer[\lrule{seq}]
	{ \ilsiljudgment{\ilderiv{\ilderiv{d_1}{P}{\regr_1}{R} & \ilderiv{d_2}{R}{\regr_2}{Q}}{P}{\regr_1; \regr_2}{Q}}{P'}{\regr_1; \regr_2}{Q'} }
	{ \ilsiljudgment{\ilderiv{d_1}{P}{\regr_1}{R}}{P'}{\regr_1}{R'} & \ilsiljudgment{\ilderiv{d_2}{R}{\regr_2}{Q}}{R'}{\regr_2}{Q'} }
	\]

	\proofcase{\ilrule{er\mbox{-}id}}
	By hypothesis, $Q' \implies Q = P$. Therefore, if we take $P' = Q' = \code{UTurn d Q'}$ we have that the triple is provable using \lrule{er\mbox{-}id}.

	\proofcase{\ilrule{choiceL}, \ilrule{choiceR}}
	We focus on the case for \ilrule{choiceL}, \ilrule{choiceR} is analogous.
	The hypothesis of the IL triple is that $\iltriple{P}{\regr_1}{Q}$ is provable with some derivation $d$. By inductive hypothesis, we then get a $P' = \code{UTurn d Q'}$ such that $\ilsiljudgment{\ilderiv{d}{P}{\regr_1}{Q}}{P'}{\regr_1}{Q'}$. Since by hypothesis $Q' \nequiv \false$, by soundness (Theorem~\ref{th:uturn-soundness}) also $\false \nequiv P' \implies P$. We can conclude the proof of the desired U-Turn judgment with rule \lrule{choiceL}.

	\proofcase{\ilrule{iter0}}
	By hypothesis, $Q' \implies Q = P$. Therefore, if we take $P' = Q' = \code{UTurn d Q'}$ we have that the triple is provable using \lrule{iter0}.

	\proofcase{\ilrule{unroll}}
	The hypothesis of the IL triple is that $\iltriple{P}{\regr^{\kstar}; \regr}{Q}$ is provable with some derivation $d$. By inductive hypothesis, we then get a $P' = \code{UTurn d Q'}$ such that $\ilsiljudgment{\ilderiv{d}{P}{\regr^{\kstar}; \regr}{Q}}{P'}{\regr^{\kstar}; \regr}{Q'}$. Since by hypothesis $Q' \nequiv \false$, by soundness (Theorem~\ref{th:uturn-soundness}) also $\false \nequiv P' \implies P$. We can conclude the proof of the desired U-Turn judgment with rule \lrule{unroll}.
\end{proof}

\begin{proof}[Proof of Theorem~\ref{th:uturn-il-valid}]
	The proof is by induction on the derivation tree. We recall that, by soundness of the IL proof system, it is enough to show $\vdash \iltriple{P'}{\regr}{Q'}$.
	
	Cases \lrule{assign} and \lrule{nondet} are Lemma~\ref{lmm:assign-valid-for-il} and \ref{lmm:nondet-valid-for-il}, respectively.
	
	Cases \lrule{skip}, \lrule{disj}, \lrule{seq}, \lrule{er\mbox{-}id}, \lrule{choiceL}, \lrule{choiceR}, \lrule{iter0} and \lrule{unroll} are all proved by using the inductive hypothesis to obtain that the premises are valid IL triples, and then prove the consequence by using the homonymous IL rule.
	
	\proofcase{\lrule{assume}}
	Since $Q' \implies P \land b$, in particular $Q' \implies b$, so $Q' \land b \equiv Q'$. We can then prove the IL triple $\vdash \iltriple{Q'}{\code{b?}}{Q'}$ via \ilrule{assume} by taking $P = Q'$ and using the previous equivalence.
	
	\proofcase{\lrule{empty}}
	$\iltriple{\false}{\regr}{\false}$ can be proved with IL rule \ilrule{empty}.
	
	\proofcase{\lrule{consIL}}
	Since the triple $\siltriple{P''}{\regr}{Q''}$ is the same in the premise and in the conclusion, the thesis is exactly the inductive hypothesis.
\end{proof}

\end{document}